\documentclass[showpacs,showkeys,12pt,preprint,preprintnumbers,nofootinbib,groupedaddress,superscriptaddress,amsmath,amssymb]{revtex4}

\usepackage[dvips]{color}
\usepackage{graphicx}
\usepackage{epsfig}    
\usepackage{dcolumn}
\usepackage{bm}
\usepackage{here}

\def\Journal#1#2#3#4{{#1} {\bf #2}, #3 (#4)}

\def\NPB{{\em Nucl. Phys.} B}
\def\PLB{{\em Phys. Lett.}  B}
\def\PRL{\em Phys. Rev. Lett.}
 
\def\PRD{{\em Phys. Rev.} D}

\def\EPC{{\em Euro. Phys. J.} C}

\allowdisplaybreaks[4]

\begin{document}

\title{
Lepton flavor violation in Higgs boson decays under the rare tau 
decay results
}

\author{Shinya Kanemura}
\email{kanemu@het.phys.sci.osaka-u.ac.jp}
\affiliation{Department of Physics, Osaka University, Toyonaka, Osaka
 560-0043, Japan}

\author{Toshihiko Ota}
\email{toshi@het.phys.sci.osaka-u.ac.jp}
\affiliation{Department of Physics, Osaka University, Toyonaka, Osaka
 560-0043, Japan}

\author{Koji Tsumura}
\email{ko2@het.phys.sci.osaka-u.ac.jp}   
\affiliation{Department of Physics, Osaka University, Toyonaka, Osaka
 560-0043, Japan}

\preprint{OU-HET 521}



\begin{abstract}

We study lepton flavor violation (LFV) associated with 
tau leptons in the framework of the two Higgs doublet model, 
in which LFV couplings are introduced as a deviation from 
Model II Yukawa interaction.
Parameters of the model are constrained from experimental results 
and also from requirements of theoretical consistencies such as 
vacuum stability and perturbative unitarity.
Current data for rare tau decays provide substantial upper limits    
on the LFV Yukawa couplings in the large $\tan\beta$ region 
($\tan\beta \gtrsim 30 $), which are comparable with 
predictions in fundamental theories. 
Here $\tan\beta$ is the ratio of vacuum expectation values 
of the two Higgs doublets.
We show that a search for the LFV decays 
$\phi^{0} \rightarrow \tau^\pm \mu^\mp$ $(\tau^\pm e^\mp)$ 
of neutral Higgs bosons ($\phi^{0} =h,H$ and $A$) 
at future collider experiments
can be useful to further constrain the LFV couplings 
especially in the relatively small $\tan\beta$ region 
($\tan\beta \lesssim 30 $),  
where rare tau decay data cannot give any strong limit.
\end{abstract}

\maketitle

\section{Introduction}

Experimental determination of the electroweak symmetry breaking sector
is important not only to confirm 
the Higgs mechanism and the mass generation mechanism for matter
but also to obtain information for physics beyond the 
standard model (SM).
Lots of new physics models predict extended Higgs sectors
with more than one scalar doublets 
in the low energy effective theories. 
Such extended Higgs sectors would show distinctive features from the SM
phenomenology. The most obvious evidence is the confirmation of
the existence of the extra scalar
states such as CP-odd and charged states.
Even when they are too heavy to be directly detected and only the
lightest Higgs boson is found at experiments, we can explore a 
possibility of the extended Higgs sector by looking for  
deviations from the SM predictions in its couplings with gauge 
bosons and fermions as well as in the self coupling. 
Moreover, it can also be examined by searching for non-SM interactions. 

Lepton flavor violation (LFV) is an example for such non-SM phenomena. 
In particular, LFV in the Yukawa sector can only appear 
for extended Higgs sectors. 
Flavor violation between electrons and muons\cite{KunoOkada}   
has been tested through rare muon decays such as  
$\mu^{}\rightarrow e^{} \gamma$ and  
$\mu^{} \rightarrow e^{}e^{+}e^{-}$,  
as well as through $\mu$-$e$ conversion.  
Tau lepton associated LFV has also been studied by rare decays of 
tau leptons such as 
$\tau \rightarrow \ell_i P^0 $\cite{belle-tau-lp0},
$\tau \rightarrow \ell_i M^+M'^-$\cite{belle-tau-lMM,babar-tau-lMM}, 
$\tau \rightarrow \ell_i\ell'^+\ell'^-$\cite{belle-tau-3l,babar-tau-3l}, and   
$\tau \rightarrow \ell_i\gamma$\cite{belle-tau-egamma,
belle-tau-mugamma,babar-tau-mugamma},
where $\ell_i$ ($i=1,2$) respectively represent an electron and a muon, 
$P^0$ does $\pi^0$, $\eta$ and $\eta'$ mesons, 
$M^\pm$ ($M'^\pm$) does $\pi^\pm$ and $K^\pm$ mesons, 
and $\ell'^\pm=e^\pm$ and $\mu^\pm$.
The LFV Yukawa couplings can be constrained from 
the data for these processes especially those with the Higgs boson mediation. 
For $\mu$-$e$ mixing, the Higgs boson mediated LFV coupling 
has been discussed in Ref.~\cite{KitanoKoikeKomineOkada,muegamma-type3THDM}. 
Tau lepton associated LFV processes with the Higgs boson mediation 
have been discussed in models with 
supersymmetry (SUSY)\cite{BK,DER,Sher-tmeta,Rossi}
as well as in the two Higgs doublet model (THDM) 
in some specific scenarios\cite{cheng-sher,HiggsLFV-THDM,Iltan}.
In Ref.~\cite{BHHS}, tau associated LFV processes have 
been discussed comprehensively 
in the framework of 4-Fermi contact interactions. 
Phenomenological consequences of the LFV Yukawa couplings 
associated with tau leptons have also been studied 
for future observables 
such as $B_s$ decays\cite{bsmutau,DER} at 
(super) B factories\cite{super-B} and 
Higgs boson decays\cite{Pilaftsis,Rossi,Herrero} at CERN LHC\cite{Assamagan}, 
an electron-positron linear collider (LC)\cite{Osaka} 
and a muon collider\cite{muon-collider}.
In addition, it has been pointed out that 
deep inelastic scattering processes 
$\mu N \to \tau X$\cite{Sher-turan,mutauDIS} from 
intense high energy muons at neutrino factories (or muon colliders)
and $e N \to \tau X$\cite{mutauDIS} by using the electron (positron) 
beam of a LC would be useful to further explore 
the tau lepton associated LFV Yukawa couplings. 

In this paper, we study LFV in Higgs boson decays  
into a $\tau$-$\ell_i$ pair in the general framework of the THDM. 
In order to evaluate possible maximal values of the branching fractions, 
we first study experimental upper limits 
on the tau lepton associated LFV Yukawa couplings. 
The parameter space is 
tested by theoretical requirements for vacuum stability\cite{VS} 
and perturbative unitarity\cite{PU-1,PU-2,PU-3}.
Current data from electroweak precision measurements 
at LEP\cite{LEP, rho-param} and those at the B factories\cite{bsg-ex, bsg} 
also strongly constrain parameters of the Higgs potential. 
Under these theoretical bounds and experimental limits on the model, 
possible maximal values of the
LFV couplings of $\tau$-$\ell_i$-$\phi^{0}$ are obtained 
by using the current data for rare tau decays, where 
$\phi^{0}$ represents two CP-even ($h$ and $H$) 
and a CP-odd ($A$) Higgs bosons. 
We then evaluate branching ratios of $\phi^0 \to \tau^\pm \ell_i^\mp$ 
with the maximal allowed values of the 
LFV couplings of $\tau$-$\ell_i$-$\phi^{0}$ 
in a wide range of the parameter space. 

We here consider the model in which Higgs self-interactions 
and quark Yukawa interactions are invariant under
the discrete symmetry ($\Phi_{1} \rightarrow + \Phi_{1}$ and $\Phi_{2}
\rightarrow - \Phi_{2}$ with $\Phi_{a}$ ($a=1,2$) being the Higgs
doublets) 
in order to suppress flavor changing neutral current (FCNC)\cite{discretesym}.
If the discrete symmetry is exact, only two choices are possible for 
Yukawa interaction; i.e., so called Model I and Model II\cite{HHG}. 
In Model I only one Higgs doublet gives masses of quarks and leptons, 
while one of the Higgs doublets gives masses of up-type quarks and the
other does of down-type quarks and charged leptons in Model II. 
We assume the Model II Yukawa interaction for quarks assigning 
$q_L^i \to + q_L^i$, $u_R^i \to - u_R^i$ and $d_R^i \to + d_R^i$. 
Even in Model II (or Model I), FCNC can be induced 
at one loop level due to off-diagonal elements of 
the Cabibbo-Kobayasi-Maskawa (CKM) matrix, and in case due to 
new physics effects. 
Flavor non-conserving decays of 
Higgs bosons into quark pairs have been studied in the 
THDM\cite{cheng-sher,qfv-hdecay-thdm} and also 
in the context of the SUSY models\cite{qfv-hdecay-susy}. We do not 
discuss flavor violation in the quark sector in this paper. 
For leptonic Yukawa interactions, the discrete symmetry is 
assumed to be explicitly broken, so that the LFV Yukawa couplings 
naturally appear in the model. 
Once a fundamental model is specified at high energy scales, 
the LFV Yukawa couplings are predicted in terms of the model parameters.
In the minimal supersymmetric standard model (MSSM)
slepton mixing may be the origin of 
LFV\cite{MSSMRN,BK,DER}.
Lepton flavor violating interactions are also induced at the loop level 
in the Zee model where tiny neutrino masses are explained 
by the dynamics of the extended Higgs sector\cite{Zee}.
We here do not specify the origin of LFV, and 
treat the lepton flavor violating THDM as 
the low energy effective theory of such high energy theories. 

It turns out that rare tau decay searches can give   
substantial upper limits on the LFV couplings of 
$\tau$-$\ell_i$-$\phi^{0}$ 
in the large $\tan\beta$ region, which are comparable with 
the values predicted by assuming some fundamental models 
beyond the THDM. 
Here $\tan\beta$ is the ratio of vacuum expectation values 
for neutral components of the Higgs doublets. 
The upper limits are rapidly relaxed for smaller $\tan\beta$ values. 
Therefore, under the constraint from current experimental data 
and the theoretical bounds, sufficiently large branching ratios 
of $\phi^0 \to \tau^\pm \ell_i^\mp$ are possible except for 
extremely large $\tan\beta$ values, which can be tested 
at future collider experiments. 
We conclude that a search for these LFV Higgs boson decays 
can be useful to explore the LFV couplings especially in 
the parameter region where rare tau decay results cannot reach.
In particular, the LFV decays of the lightest Higgs boson 
can be one of the important probes for extended Higgs sectors 
when the SM-like situation would be preferred by the data 
at the experiments.

In Sec.~\ref{Sec:Model}, the THDM with LFV 
in Yukawa interaction is defined. 
The allowed region of the parameters of the Higgs sector 
is discussed by requiring the theoretical consistencies 
and the experimental results.
In Sec.~\ref{Sec:LFV-tau-decay},  
we discuss the upper limit on the LFV Yukawa couplings 
in a wide range of the parameter space  
by using the current data for the rare tau decays.
In Sec.~\ref{Sec:LFV-Hdecay}, LFV in Higgs boson decays 
is studied under the rare tau decay results.
Conclusions are given in Sec.~\ref{Sec:conclusion}. 

\section{Model}
\label{Sec:Model}

In this section, the lepton flavor violating THDM is introduced, 
and the theoretical constraints and the experimental limits 
are discussed.

\subsection{Lagrangian}

We consider the Yukawa interaction for charged leptons,  
\begin{align}
-\mathcal{L}_{\text{lepton}}=
\overline{\ell}_{R i} 
 \left\{
 Y_{\ell_{i}} \delta_{ij} \Phi_{1}  
 +
 \left(
 Y_{\ell_{i}} \epsilon^{L}_{ij} 
 +
 \epsilon^{R}_{ij} Y_{\ell_{j}}
 \right) \Phi_{2} 
 \right\}
 \cdot
 L_{j} + {\rm H.c.},
\label{eq:Llepton}
\end{align}
where 
$\Phi_{1}$ and $\Phi_{2}$ are the scalar iso-doublets with hypercharge $1/2$,
$\ell_{R i}$ ($i$=1-3) are singlet fields for right-handed charged leptons,
$L_{i}$ ($i$=1-3) denote the lepton doublets 
and $Y_{\ell_{i}} $ are the Yukawa
couplings for $\ell_{i}$. 
This interaction is reduced to be of Model II\cite{HHG}  
in the limit $\epsilon^{L,R}_{ij} \rightarrow 0$ with 
the discrete symmetry under $e_{R}^{i} \rightarrow + e_{R}^{i}$,
$L_{i} \rightarrow + L_i$,
$\Phi_{1} \rightarrow + \Phi_{1}$, and
$\Phi_{2} \rightarrow - \Phi_{2}$.   
Nonzero values of $\epsilon_{ij}^{L,R}$ ($i\neq j$)
yield the LFV Yukawa couplings after the diagonalization 
of the mass matrix.
We note that in supersymmetric standard models, 
the Yukawa interaction for leptons is of Model II 
at the tree level, and $\epsilon_{ij}^{L,R}$ 
can be induced at the loop level due to slepton 
mixing\cite{MSSMRN,BK,DER}.
For the quark sector, Model II Yukawa interactions are assumed to 
suppress FCNC, 
imposing the invariance under the transformation of 
$u_{R}^{i} \rightarrow - u_{R}^{i}$, $d_{R}^{i} \rightarrow + d_{R}^{i}$,
$q_{L}^{i} \rightarrow + q_{L}^{i}$,
$\Phi_{1} \rightarrow + \Phi_{1}$, and
$\Phi_{2} \rightarrow -\Phi_{2}$.

The Higgs sector of the general THDM is expressed as
\begin{align}
-\mathcal{L}_{\text{Higgs}} 
=&
m_{1}^{2} \left| \Phi_{1} \right|^{2}
+
m_{2}^{2} \left| \Phi_{2} \right|^{2}
-
\left( m_{3}^{2} \Phi_{1}^{\dagger} \Phi_{2} + {\rm H.c.} \right)
+
\frac{\lambda_{1}}{2} \left|\Phi_{1}\right|^{4}
+
\frac{\lambda_{2}}{2} \left|\Phi_{2}\right|^{4} \nonumber \\
&+
\lambda_{3} \left|\Phi_{1}\right|^{2} \left|\Phi_{2}\right|^{2}
+
\lambda_{4} \left|\Phi_{1}^{\dagger} \Phi_{2} \right|^{2}
+
\left\{
 \frac{\lambda_{5}}{2}
 \left( \Phi_{1}^{\dagger} \Phi_{2} \right)^{2} + {\rm H.c.}
\right\} \nonumber \\
&+ \left(
 \lambda_{6} \left|\Phi_{1}\right|^{2}
 \Phi_{1}^{\dagger}
 \Phi_{2} +{\rm H.c.} \right) 
 + \left(
 \lambda_{7} \left|\Phi_{2}\right|^{2}
 \Phi_{1}^{\dagger}
 \Phi_{2} +{\rm H.c.} \right).
\label{eq:LHiggs}
\end{align}
In Eq.~\eqref{eq:LHiggs}, 
$m_{3}^{2}$, $\lambda_{5}$, $\lambda_{6}$ and
$\lambda_{7}$ are complex in general.
We here assume that all the parameters $m_{1-3}^{2}$ and $\lambda_{1-7}$ are
real. 
The terms of $m_{3}^{2}$, $\lambda_{6}$ and
$\lambda_{7}$ break the discrete symmetry explicitly. 
As we consider the model in which the discrete symmetry is explicitly 
broken only in the leptonic Yukawa interaction, 
we set the hard-breaking coupling constants 
to be zero in the Higgs potential; 
i.e., $\lambda_{6} = \lambda_{7} =0$\footnote{%
Even in such a case, $\lambda_{6}$ and $\lambda_{7}$
are effectively induced at the loop level.
They are suppressed by the loop factor, so that we here neglect
these small effects.}, 
and retain only the soft-breaking mass parameter $m_3^2$.

There are eight degrees of freedom in the two Higgs doublet fields.
Three of them are absorbed by the weak gauge bosons via the Higgs
mechanism. Remaining five are physical states.
After the diagonalization of the mass matrices, they correspond to 
two CP-even ($h$ and $H$), 
a CP-odd ($A$), and a pair of charged  ($H^{\pm}$) Higgs bosons.
The CP-even neutral states are defined such that $h$ is lighter than $H$.
In our model, 
the eight real parameters $m_{1-3}^{2}$ and $\lambda_{1-5}$ 
can be described by the same number of physical parameters; i.e.,
the vacuum expectation value $v$ $(\simeq 246$ GeV), 
the Higgs boson masses 
$m_{h}^{}, m_{H}^{}, m_{A}^{}$ and  $m_{H^{\pm}}^{}$,  
the mixing angle $\alpha$ between the CP-even Higgs bosons, 
the ratio $\tan \beta$ 
($ \equiv \langle \Phi^{0}_{2} \rangle / \langle \Phi^{0}_{1}
\rangle$) 
of the vacuum expectation values for two Higgs doublets,
and the soft-breaking scale $M$ ($\equiv \sqrt{m_{3}^{2}/\sin\beta
\cos\beta}$) for the discrete symmetry.
The quartic couplings are expressed in terms of physical parameters 
by\cite{smlike2} 
\begin{align}
\lambda_{1} =&
 \frac{1}{v^{2} \cos^{2} \beta}
 \left(
 -\sin^{2} \beta M^{2}
 +\sin^{2} \alpha m_{h}^{2}
 +\cos^{2} \alpha m_{H}^{2}
 \right), \\
\lambda_{2} =&
 \frac{1}{v^{2} \sin^{2} \beta}
 \left(
 -\cos^{2} \beta M^{2}
 +\cos^{2} \alpha m_{h}^{2}
 +\sin^{2} \alpha m_{H}^{2}
 \right), \\
\lambda_{3} =&
 \frac{1}{v^{2}}
 \left\{
 - M^{2}
 +2 m_{H^{\pm}}^{2}
 +\frac{\sin 2 \alpha}{\sin 2 \beta} \left( m_{H}^{2} - m_{h}^{2} \right)
 \right\}, \\
\lambda_{4}
 =&
 \frac{1}{v^{2}}
 \left(
 M^{2} + m_{A}^{2} - 2 m_{H^{\pm}}^{2} 
 \right),\\
\lambda_{5}
 =&
 \frac{1}{v^{2}} \left(M^{2} - m_{A}^{2} \right).
\end{align}

The relative size of the parameter $M$ against $v$ determines the
decoupling property of the Higgs sector.
For the case of $M^{2} \gg v^{2}$, 
$m_H^{}$, $m_A^{}$ and $m_{H^{\pm}}$ turn out to be 
around $M$. 
They decouple from low energy observables by a factor of $v^{2}/M^{2}$,
and the lightest Higgs boson $h$ becomes the SM-like one\cite{GH,smlike2}.
The THDM effectively becomes the SM at low energies.
On the contrary, when $M^{2} \lesssim v^{2}$, the Higgs boson masses
can be varied by the variation of 
relative size of quartic couplings.
In such a case, the parameter space is strongly constrained by the
conditions from perturbative unitarity, which we shall discuss in the
next subsection. 
Notice that even for $M^2 \lesssim v^2$ we can define the ``SM-like'' 
limit by setting $\sin(\alpha-\beta) \sim -1$\cite{smlike1,smlike2}. 
The coupling constants of $hVV$ ($VV=W^+W^-$ and $ZZ$) 
then coincide with those in the SM at the tree level, and 
there are no $HVV$ couplings.

\subsection{Constraint on the Higgs parameters}

Parameters of the Higgs sector 
are constrained from 
requirements of theoretical consistencies 
and also from the current experimental results. 
We here take into account two kinds of theoretical conditions; i.e.,  
vacuum stability\cite{VS} and perturbative unitarity\cite{PU-1,
PU-2,PU-3} at the tree level.
The condition for vacuum stability is expressed by
\begin{align}
\lambda_{1} > 0, \quad
\lambda_{2} > 0, \quad
\sqrt{\lambda_{1} \lambda_{2}} +\lambda_{3} 
+ {\rm MIN} 
\left[ 0, \lambda_{4} + \lambda_{5} , \lambda_{4}-\lambda_{5}
 \right] >0, 
\label{eq:VS}
\end{align}
and that for tree-level unitarity is described as
\begin{align}
\left|\langle \phi_{3} \phi_{4}  | a^{0}  | \phi_{1} \phi_{2} \rangle   \right|
< \xi, 
\label{eq:PU}
\end{align}
where
$\langle \phi_{3} \phi_{4}  |a^{0}  | \phi_{1} \phi_{2} \rangle $ is the 
$s$-wave
amplitude for the process of $\phi_{1} \phi_{2} \rightarrow \phi_{3}
\phi_{4}$ with 
$\phi_{a}$ ($a$=1-4) denoting Higgs bosons 
and longitudinal components of weak gauge bosons.
We employ the conditions given in Refs.~\cite{PU-2,smlike2} in which 
the fourteen channels are taken into account in the THDM; i.e.,
$ZZ$, $hZ$, $HZ$, $AZ$, $hh$, $HH$, $AA$, $hH$, $hA$, $HA$, 
$W^+W^-$, $H^+H^-$ and $W^\pm H^\mp$. 
We take the criterion $\xi$ 
to be 1\cite{PU-1} (and also 1/2\cite{HHG} for comparison).
For the top and bottom Yukawa couplings with Higgs bosons 
($y_{t\bar{t}\phi^{0}}$
and $y_{b\bar{b}\phi^{0}}$), we just put the criterion,  
\begin{align}
|y_{q\bar{q}\phi^{0}}|^{2} < 4 \pi, \quad (q=t,b).
\label{eq:yukawa-bound}
\end{align}
For example, this condition on $y_{t\bar{t}A}$ and 
$y_{b\bar{b}A}$ gives upper and lower limits for $\tan\beta$ 
($0.2 \lesssim \tan\beta \lesssim 100$-$200$)\footnote{
Similar bounds are obtained for example in Ref.~\cite{vbarger}.}. 
As shown in the next section,
the conditions of Eqs.~\eqref{eq:VS} and \eqref{eq:PU}
give more strict upper bound on $\tan\beta$ 
than that of Eq.~\eqref{eq:yukawa-bound} 
in a wide parameter region.

Next we consider the experimental constraints, which are provided by the
LEP precision data\cite{LEP}, 
the $b \rightarrow s \gamma$ results\cite{bsg-ex}, 
and the direct search results for the Higgs bosons\cite{LEP}.  
The LEP precision data provide the strong constraint 
on the new physics structure via the gauge-boson two-point functions. 
The constraint on $\rho$ parameter indicates that the Higgs sector 
is approximately custodial $SU(2)$ symmetric.
This requirement is satisfied when 
(i) $m_{H^{\pm}}^{} \simeq m_{A}^{}$,  
(ii) $m_{H^{\pm}}^{} \simeq m_{H}^{}$ with
$\sin^{2} (\alpha - \beta)\simeq 1$, and 
(iii) $m_{H^{\pm}}^{} \simeq m_{h}^{}$ with
$\cos^{2} (\alpha - \beta) \simeq 1$\cite{rho-param}.
It is known that in Model II,  the $b \rightarrow s \gamma $ result 
gives the lower bound on the charged Higgs boson mass\cite{bsg,bsg-ex}.
We here take into account this bound by requiring
$m_{H^{\pm}}^{} \gtrsim 350$ GeV.

\subsection{LFV parameters $\kappa^{L,R}_{ij}$}

The tau lepton associated LFV interactions 
in Eq.~\eqref{eq:Llepton} can be reduced 
in the mass eigenbasis of each field to
\begin{align}
-\mathcal{L}_{\tau{\rm LFV}}
=&
\frac{m_{\tau}}{v \cos^{2}\beta}
 \left(
 \kappa^{L}_{3i} \overline{\tau} {\rm P}_{L} \ell_{i}
 +
 \kappa^{R}_{i3} \overline{\ell}_i {\rm P}_{L} \tau
 \right)
 \left\{
 \cos\left(\alpha - \beta \right) h
 +
 \sin\left(\alpha - \beta \right) H
 -
 {\rm i} A
 \right\} \nonumber \\
 &+
 \frac{\sqrt{2} m_{\tau}}{v \cos^{2}\beta}
 \left(
 \kappa^{L}_{3i}
 \overline{\tau} {\rm P}_{L} \nu_{i} 
 +
 \kappa^{R}_{i3}
 \overline{\ell_{i}} {\rm P}_{L} \nu_{\tau}
 \right)  H^{-}
 +
 {\rm H.c.},
\label{eq:tauLFV}
\end{align}
where ${\rm P}_{L}$ is the projection operator to 
the left-handed field, and 
$\ell_1$ and $\ell_2$ respectively represent $e$ and $\mu$.
In general, the LFV parameters $\kappa^{L,R}_{ij}$ can be expressed in
terms of $\epsilon^{L,R}_{ij}$ and $\tan\beta$.\footnote{%
LFV parameters $\kappa_{ij}^{L,R}$ can be expressed as 
\begin{align}
\kappa^{X}_{ij} 
= 
-\frac{\epsilon^{X}_{ij}}{
 \left\{ 1+ (\epsilon^{L}_{33} + \epsilon^{R}_{33})\tan\beta
 \right\}^{2} }
\qquad (X=L,R),
\end{align}
for the case of 
$\epsilon^{L,R}_{ij} \tan\beta \ll \mathcal{O}(1)$ 
which is satisfied in the MSSM 
case\cite{BK,DER}.}
We here take these $\kappa^{L,R}_{ij}$ as effective couplings,
and investigate their phenomenological consequences.
We note that Eq.~(\ref{eq:tauLFV}) is exact in the limit of 
$m_{\ell_{i}}^{} \to 0$.
The terms of $\kappa_{i3}^L$ and $\kappa_{3i}^R$ ($i=1, 2$) 
are proportional to $m_{\ell_i}^{}$, 
so that they decouple in this approximation.

We briefly discuss relationship between 
$\kappa^{L,R}_{ij}$ and new physics models beyond the
cut-off scale of the {\it effective} THDM.
When a new physics model is specified at the high energy scale, 
$\kappa^{L,R}_{ij}$ can be predicted as a function of the model 
parameters.
For example, in the MSSM, slepton mixing can be a source of LFV.
Notice that the induced LFV Higgs interactions do not necessarily 
decouple in the limit where the SUSY  particles are sufficiently heavy, 
because their couplings only depend on the ratio of the SUSY 
parameters.
Therefore, the Higgs associated LFV processes can become
important in a scenario with the soft-SUSY-breaking scale 
$m_{\rm SUSY}^{}$ to be much higher than the electroweak one. 
In the MSSM, predicted values of $|\kappa_{3i}^L|^2$ 
can be as large as of ${\cal O}({10^{-6}})$ when 
$m_{\rm SUSY}^{}$ is a few TeV\cite{Rossi,Osaka}.
In the MSSM with right-handed neutrinos, 
slepton mixing may be a consequence of running effects of the
neutrino Yukawa couplings between the scale of the grand unification 
and that of the right-handed neutrinos\cite{MSSMRN}.
The parameters $\kappa^{L}_{3i}$ are mainly induced by mixing 
of left-handed sleptons\cite{BK,DER,Sher-tmeta,Rossi,Osaka,Herrero}.
The LFV Yukawa interactions can also appear effectively in the Zee 
model\cite{Zee}.
The LFV parameters $\kappa^{L,R}_{ij}$ are induced through flavor
violating couplings in the charged scalar interactions with leptons.

\begin{table}[t] {
\begin{tabular}{lll} 
\hline \hline
Mode &    Belle (90\% CL) \hspace*{1cm}& BaBar (90\% CL)  \\
\hline 
$\tau^-\rightarrow e^-\pi^0$&
\underline{$1.9\times 10^{-7}$}\cite{belle-tau-lp0}&  \\
$\tau^-\rightarrow e^-\eta$&
\underline{$2.4\times 10^{-7}$}\cite{belle-tau-lp0}&  \\
$\tau^-\rightarrow e^-\eta'$&
\underline{$10\times 10^{-7}$}\cite{belle-tau-lp0}&  \\
$\tau^-\rightarrow \mu^-\pi^0$&
\underline{$4.1\times 10^{-7}$}\cite{belle-tau-lp0}&  \\
$\tau^-\rightarrow \mu^-\eta$&
\underline{$1.5\times 10^{-7}$}\cite{belle-tau-lp0}&  \\
$\tau^-\rightarrow \mu^-\eta'$&
\underline{$4.7\times 10^{-7}$}\cite{belle-tau-lp0}&  \\
\hline
$\tau^-\rightarrow e^-\pi^+\pi^-$&
$8.4\times 10^{-7}$\cite{belle-tau-lMM}&
\underline{$1.2\times 10^{-7}$}\cite{babar-tau-lMM}  \\
$\tau^-\rightarrow e^-\pi^+K^-$&
$5.7\times 10^{-7}$\cite{belle-tau-lMM}&
\underline{$3.2\times 10^{-7}$}\cite{babar-tau-lMM}  \\
$\tau^-\rightarrow e^-K^+\pi^-$&
$5.6\times 10^{-7}$\cite{belle-tau-lMM}&
\underline{$1.7\times 10^{-7}$}\cite{babar-tau-lMM}  \\
$\tau^-\rightarrow e^-K^+K^-$&
$3.0\times 10^{-7}$\cite{belle-tau-lMM}&
\underline{$1.4\times 10^{-7}$}\cite{babar-tau-lMM}  \\
$\tau^-\rightarrow \mu^-\pi^+\pi^-$&
\underline{$2.8\times 10^{-7}$}\cite{belle-tau-lMM}&
$2.9\times 10^{-7}$\cite{babar-tau-lMM}  \\
$\tau^-\rightarrow \mu^-\pi^+K^-$&
$6.3\times 10^{-7}$\cite{belle-tau-lMM}&
\underline{$2.6\times 10^{-7}$}\cite{babar-tau-lMM}  \\
$\tau^-\rightarrow \mu^-K^+\pi^-$&
$15.5\times 10^{-7}$\cite{belle-tau-lMM}&
\underline{$3.2\times 10^{-7}$}\cite{babar-tau-lMM}  \\
$\tau^-\rightarrow \mu^-K^+K^-$&
$11.7\times 10^{-7}$\cite{belle-tau-lMM}&
\underline{$2.5\times 10^{-7}$}\cite{babar-tau-lMM}  \\
\hline
$\tau^-\rightarrow e^-e^+e^-$\hspace*{1cm}&
$3.5\times10^{-7}$\cite{belle-tau-3l}&
\underline{$2.0\times10^{-7}$}\cite{babar-tau-3l}  \\
$\tau^-\rightarrow e^-\mu^+\mu^-$&
\underline{$2.0\times 10^{-7}$}\cite{belle-tau-3l}&
$3.3\times 10^{-7}$\cite{babar-tau-3l}  \\
$\tau^-\rightarrow \mu^-e^+e^-$&
\underline{$1.9\times 10^{-7}$}\cite{belle-tau-3l}&
$2.7\times 10^{-7}$\cite{babar-tau-3l}  \\
$\tau^-\rightarrow \mu^-\mu^+\mu^-$&
$2.0\times 10^{-7}$\cite{belle-tau-3l}&
\underline{$1.9\times 10^{-7}$}\cite{babar-tau-3l}  \\
\hline
$\tau\rightarrow e\gamma$&
\underline{$3.9\times 10^{-7}$}\cite{belle-tau-egamma}&  \\
$\tau\rightarrow \mu\gamma$&
$3.1\times 10^{-7}$\cite{belle-tau-mugamma}& 
\underline{$6.8\times 10^{-8}$}\cite{babar-tau-mugamma}   \\
\hline \hline
\end{tabular}}
\caption{Current experimental limits 
on branching ratios of the LFV rare tau decays.}
\label{Tab:tau-bound}
\end{table}

\section{Bound on LFV Yukawa couplings from
 rare tau decays} 
\label{Sec:LFV-tau-decay}

In order to constrain the LFV parameters 
$|\kappa_{3i}^{L}|$ and $|\kappa_{i3}^{R}|$,
we take into account the data for rare tau decay processes 
such as 
$\tau \rightarrow \ell_i P^0 $,
$\tau \rightarrow \ell_i M^+M'^-$, 
$\tau \rightarrow \ell_i\ell'^+\ell'^-$, and   
$\tau \rightarrow \ell_i\gamma$,
where $P^0$ represents $\pi^0$, $\eta$ and $\eta'$ mesons, 
$M^\pm$ ($M'^\pm$) does $\pi^\pm$ and $K^\pm$ mesons, 
and $\ell'^\pm=e^\pm$ and $\mu^\pm$.
The list of the current data from the B factories are 
summarized in Table~\ref{Tab:tau-bound}
\cite{belle-tau-lp0,belle-tau-lMM,babar-tau-lMM,
belle-tau-3l,babar-tau-3l,belle-tau-egamma,
belle-tau-mugamma,babar-tau-mugamma}. 
These bounds may be improved at the super B factory
around a digit\cite{super-B}. 
In our analysis, we take the underlined data in
Table~\ref{Tab:tau-bound} as our numerical inputs. 
The branching ratios for these $\tau^{-}$ decays are 
calculated in our model as
\begin{align}
&{\rm Br}(\tau^{-} \rightarrow \ell^{-}_i \pi^0) = 
 \frac{3 G_{F}^{2} m_{\tau}^{3} m_{\pi^0}^{4} F_{\pi}^{2}
 \tau_{\tau}\left|   \kappa_{3i}
 \right|^{2} }{32\pi \cos^{4} \beta}
 \left(1-\frac{m_{\pi^0}^{2}}{m_{\tau}^{2}}\right)^{2}
 \left(\frac{m_u \cot\beta - m_d \tan\beta}
{m_u+m_d}\right)^2 \frac{1}{m_{A}^{4}}, 
\label{eq:tmpi0}\\
&{\rm Br}(\tau^{-} \rightarrow \ell^{-}_i \eta) = 
 \frac{9 G_{F}^{2} m_{\tau}^{3} m_{\eta}^{4} F_{\eta}^{2}
 \tau_{\tau}\left|
   \kappa_{3i}
 \right|^{2} }{32\pi \cos^{4} \beta}
 \left(1-\frac{m_{\eta}^{2}}{m_{\tau}^{2}}\right)^{2}
 \left(\frac{m_u \cot\beta + (m_d - 2 m_s) \tan\beta}
{m_u+m_d+4 m_s}\right)^2 \frac{1}{m_{A}^{4}}, 
\label{eq:tmeta}\\
&{\rm Br}(\tau^{-} \rightarrow 
\ell^{-}_i \pi^{+} \pi^{-})  = 
 \frac{G_{F}^{2} m_{\tau}^{5} B_{0}^{2} \tau_{\tau} \left|\kappa_{3i}\right|^{2}
}{256 \pi^{3}
 \cos^{6} \beta}
 \biggl\{  m_{d} 
 \left(
  \frac{\sin\alpha \cos(\alpha-\beta)}{m_{h}^{2}}
 -
  \frac{\cos\alpha \sin(\alpha -\beta)}{m_{H}^{2}}
 \right) \nonumber \\
&\hspace{1.8cm} -
 m_{u} \cot\beta
  \left(
  \frac{\cos\alpha \cos(\alpha-\beta)}{m_{h}^{2}}
  +
  \frac{\sin\alpha \sin(\alpha-\beta)}{m_{H}^{2}}
  \right)
 \biggr\}^{2}, \label{eq:tmpp}\\
&{\rm Br}(\tau^{-} \rightarrow 
\ell^{-}_i K^{+} K^{-})  = 
 \frac{G_{F}^{2} m_{\tau}^{5} B_{0}^{2} \tau_{\tau}
 \left|\kappa_{3i}\right|^{2}}{256 \pi^{3}
 \cos^{6} \beta}
 \biggl\{  m_{s} 
 \left(
  \frac{\sin\alpha \cos(\alpha-\beta)}{m_{h}^{2}}
 -
  \frac{\cos\alpha \sin(\alpha -\beta)}{m_{H}^{2}}
 \right) \nonumber \\
&\hspace{1.8cm} -
 m_{u} \cot\beta
  \left(
  \frac{\cos\alpha \cos(\alpha-\beta)}{m_{h}^{2}}
  +
  \frac{\sin\alpha \sin(\alpha-\beta)}{m_{H}^{2}}
  \right)
 \biggr\}^{2}, \label{eq:tmkk}\\
&{\rm Br}(\tau^{-} \rightarrow 
\ell^{-}_i \pi^{+} K^{-})  =
{\rm Br}(\tau^{-} \rightarrow 
\ell^{-}_i K^{+} \pi^{-}) \sim \frac{G_F^2 m_W^4}{32 \pi^4 } 
  |V_{ud}V_{us}^\ast|^2   {\rm Br}(\tau^{-} \rightarrow 
\ell^{-}_i K^{+} K^{-}), 
\label{eq:tmkpi}\\
&{\rm Br}(\tau^{-} \rightarrow \ell^{-}_i \ell'^{+} \ell'^{-}) \nonumber\\
&  \hspace{1.5cm}= 
 \frac{G_{F}^{2} m_{\ell'}^{2} m_{\tau}^{7} \tau_{\tau} 
\left|\kappa_{3i} \right|^{2} 
}{1536 \pi^{3} \cos^{6} \beta} 
 \left\{\!\!
 \left(
 \frac{\sin\alpha \cos(\alpha-\beta)}{m_{h}^{2}}
 -
 \frac{\cos\alpha \sin(\alpha-\beta) }{m_{H}^{2}}
 \right)^{2} \!\!
 +
 \frac{\sin^{2}\beta}{m_{A}^{4}}\!
 \right\},
\label{eq:temm} \\
&{\rm Br}(\tau^{-} \rightarrow \ell^{-}_i \ell^{+}_{i} \ell^{-}_{i}) \nonumber\\
&  \hspace*{+1.5cm}= 
 \frac{G_{F}^{2} m_{\ell_{i}}^{2} m_{\tau}^{7} \tau_{\tau} 
\left|\kappa_{3i} \right|^{2} 
}{2048 \pi^{3} \cos^{6} \beta} 
 \Biggl[
 \left\{\!\!
 \left(
 \frac{\sin\alpha \cos(\alpha-\beta)}{m_{h}^{2}}
 -
 \frac{\cos\alpha \sin(\alpha-\beta) }{m_{H}^{2}}
 \right)^{2} \!\!
 +
 \frac{\sin^{2}\beta}{m_{A}^{4}}\!
 \right\} \nonumber \\
& \hspace*{+2cm}
 +
 \frac{2}{3}
  \left(
 \frac{\sin\alpha \cos(\alpha-\beta)}{m_{h}^{2}}
 -
 \frac{\cos\alpha \sin(\alpha-\beta) }{m_{H}^{2}}
 \right)
 \frac{\sin\beta}{m_{A}^{2}}\!
 \Biggr],
\label{eq:tmmm}\\
&{\rm Br}(\tau^{-} \rightarrow \ell^{-}_i \gamma) 
= \frac{\alpha_{\rm em} 
G_{F}^{2} m_{\tau}^{9} \tau_{\tau}}{72 (4\pi)^{4} \cos^{6}\beta}
\nonumber\\
& \hspace*{+0cm} 
\times \biggl\{
 \left| \kappa^{L}_{3i} \right|^{2} 
 \left(
   \frac{\sin\alpha \cos(\alpha -\beta)}{m_{h}^{2}} f_{-}(x_h)
 -
  \frac{\cos\alpha \sin(\alpha -\beta)}{m_{H}^{2}} f_{-}(x_H)
 +
  \frac{\sin \beta}{m_{A}^{2}} f_{+}(x_A)
 \right)^{2} \nonumber \\
&\hspace{0cm}+
  \left| \kappa^{R}_{i3} \right|^{2} 
 \left(
  \frac{\sin\alpha \cos(\alpha -\beta)}{m_{h}^{2}} f_{-}(x_h)
 -
  \frac{\cos\alpha \sin(\alpha -\beta)}{m_{H}^{2}} f_{-}(x_H)
 +
  \frac{\sin \beta}{m_{A}^{2}} f_{+}(x_A)
 +
 \frac{\sin \beta}{m_{H^{\pm}}^{2}}
 \right)^{2}
\biggr\}, 
\label{eq:tmg}
\end{align}
where $|\kappa_{3i}|^{2} \equiv |\kappa^{L}_{3i}|^{2} +
|\kappa^{R}_{i3}|^{2}$, 
$G_{F}$ is the Fermi constant, 
$F_{\pi}$ and $F_{\eta}$ are the decay constants of 
the pion and the $\eta$ meson which are defined as
shown in Ref.\cite{BHHS}, 
$\tau_{\tau}$ is the life time of 
the tau lepton, and $\alpha_{\rm em}$ is the fine structure constant. 
The expression in Eqs.~(\ref{eq:tmeta}) 
is deduced in the approximation of non-mixing between 
the octet and singlet states; i.e., $\eta=\eta_8$. 
In Eqs.~(\ref{eq:tmpp}) and (\ref{eq:tmkk}), $B_{0}$ is the matrix 
element for a pair production of pions, which is evaluated at the 
leading order as 
$B_{0} = m_{\pi^\pm}^{2} / \left( m_{u} + m_{d}
\right)  (= m_{K^\pm}^{2} / \left( m_{u} + m_{s}
\right))$ by using the chiral Lagrangian\cite{BHHS}. 
In Eq.~\eqref{eq:temm}, 
$\ell'$ represents either $e$ or $\mu$ but $\ell' \neq \ell_{i}$.
In Eq.~(\ref{eq:tmg}), $f_\pm(x)\equiv 1 \pm 3 (3+2 \ln x)$ and 
$x_{\phi^0}^{} \equiv m_\tau^2/m_{\phi^0}^2$ 
($\phi^0=h, H$ and $A$)\footnote{In this calculation, 
masses of neutrinos are ignored.}. 
We here omit the sub-leading contributions such as
$\mathcal{O}(m_{\ell_i}^{}/m_{\tau})$\footnote{%
The crossing terms  between $\kappa^{L}_{3i}$ and $\kappa^{R}_{i3}$
do not appear in this approximation.}. 
We also do not consider the contribution 
to $\tau\rightarrow \ell_iP^0$, $\tau\rightarrow \ell_i\ell'^{+}\ell'^{-}$
and $\tau\rightarrow \ell_i M^{+} M'^{-}$ 
from gauge boson mediated LFV diagrams  
which are induced at the loop level. 
In the SUSY-like limit ($m_{H} \simeq m_{A}^{}$), 
Eq.~\eqref{eq:tmmm} coincides with the results of
Ref.~\cite{DER}\footnote{
Eq.~\eqref{eq:tmmm} is also consistent with the expression 
in Ref.~\cite{Paradisi:2005tk}.}, 
and the ratio between Eq.\eqref{eq:tmeta} and Eq.~\eqref{eq:tmmm}  
reproduces the result shown in Ref.~\cite{Sher-tmeta}.
Our result of Eq.~\eqref{eq:temm} is consistent with Ref.~\cite{BHHS},
and Eqs.~\eqref{eq:tmpi0}, \eqref{eq:tmeta}, \eqref{eq:tmpp}, and 
\eqref{eq:tmkk} are also consistent with 
the results in Ref.~\cite{BHHS} up to the colour factor\footnote{The 
color factor is missing in the expressions in 
Ref.~\cite{BHHS}.}.
We keep the terms with $\mathcal{O}(m_{P^{0}}^{2}/m_{\tau}^{2})$ in 
Eqs.~\eqref{eq:tmpi0} and \eqref{eq:tmeta} although
we ignore those with $\mathcal{O}(m_{M}^{2}/m_{\tau}^{2})$
in the derivation of Eqs.~\eqref{eq:tmpp} and \eqref{eq:tmkk}.
The expression of Eq.~\eqref{eq:tmg} is compatible with 
Ref.\cite{Iltan}.
The branching ratios of the $\tau^{+}$ decay are the same as those of
$\tau^{-}$ in Eqs.~\eqref{eq:tmpi0}-\eqref{eq:tmg}. 

Some comments are in order. 
First, the processes $\tau \to \ell_i P^0$ ($P^0=\pi^0$, $\eta$ and $\eta'$) 
are mediated only by the CP-odd Higgs boson $A$, and 
their branching ratios strongly depend on masses of the outgoing mesons.
Because of the cancellation between the singlet and octet component in
the $\eta'$ meson, the bound on $|\kappa_{3i}|^2$ from 
$\tau \to \ell_i \eta'$ may be less important than 
that from $\tau \to \ell_i \eta$\cite{Rossi}. 
We do not include $\tau \to \ell_i \eta'$ in our analysis below.
From Table~\ref{Tab:tau-bound} and 
Eqs.~\eqref{eq:tmpi0}-\eqref{eq:tmeta}, it is expected that 
$\tau \to \ell_i \eta$ gives the most stringent upper limit 
on $|\kappa_{3i}|^2$ of these three modes.
Second, the expressions for the branching ratios of 
$\tau \to \ell_i M^+ M^-$ with $M^\pm =\pi^\pm$ and $K^\pm$ 
are the same up to the difference of down type quarks $d$ and $s$ as  
seen in Eqs.~\eqref{eq:tmpp} and \eqref{eq:tmkk}. 
These modes are mediated by the CP-even Higgs bosons $h$ and $H$.
As the mass of the $s$ quark is much greater than that of the $d$ quark 
while the experimental upper limits on them are the same order, 
$\tau \to \ell_i K^+ K^-$ gives the more stringent bounds 
on  $|\kappa_{3i}|^2$ than $\tau \to \ell_i \pi^+ \pi^-$ 
in a wide region of the parameter space.
It is also found that $\tau \to \ell_i \pi^\pm K^\mp$ 
cannot contribute to limit values of $|\kappa_{3i}|^2$.
Their branching ratios are one-loop induced and much smaller than 
those of $\tau \to \ell^-_i K^+ K^-$ 
due to the loop suppression factor and the CKM factor
$|V_{ud}V_{us}^\ast|^2$,  
while the data for these processes are the same order: see 
Eq.~\eqref{eq:tmkpi}.  
Third, in Eqs.~\eqref{eq:temm} and \eqref{eq:tmmm} 
the branching ratios of $\tau^- \to \ell^-_i \ell'^+ \ell'^-$ 
are proportional to the mass of $\ell'$. 
As the data for $\tau^- \to \ell^-_i \mu^+ \mu^-$ and 
$\tau^- \to \ell^-_i e^+ e^-$ are the same order, 
$\tau^-\to\ell^-_i \mu^+\mu^-$ provide 
much stronger upper limits on $|\kappa_{3i}|^2$ within the tree
approximation
because of the mass difference between electrons and muons\footnote{%
At the loop level,
the branching ratio of 
the $\tau^{-}\rightarrow \ell_{i}^{-} e^{+}e^{-}$ process 
can become larger than that of 
$\tau^{-}\rightarrow \ell_{i}^{-} \mu^{+}\mu^{-}$
because of the phase space dependence in the photonic penguin 
diagram\cite{EllisHisanoRaidalShimizu}.}.
The modes $\tau^- \to \ell^-_i \ell'^+ \ell'^-$ are mediated by all
neutral Higgs bosons $h$, $H$ and $A$. 
Finally, $\tau^- \to \ell^-_i\gamma$ are one-loop 
induced processes in our model. All kinds of the Higgs bosons 
are included in the one-loop diagrams.  
Notice that as shown in Eq.~\eqref{eq:tmg}  
the one loop diagram with a charged Higgs boson 
is asymmetric with respect to chirality of the tau lepton, 
as the neutrino in the diagram is left-handed.
In the following, we study the bound on $|\kappa_{3i}|^{2}$, 
and do not treat $|\kappa_{3i}^{L}|^{2}$
and $|\kappa_{i3}^{R}|^{2}$ separately.

Since branching ratios for $\tau^-\to\ell_i^-P^0$,
$\tau^-\to\ell_i^-M^+M^-$, $\tau^-\to\ell_i^-\ell'^+\ell'^-$ and 
$\tau^-\to\ell_i^-\gamma$ 
depend on different combinations of the Higgs boson masses,
independent information can be obtained for the model parameters 
by measuring each of them. When all the masses of Higgs bosons are large,
these decay processes decouple by a factor of $1/m_{\rm Higgs}^{4}$.
Although these branching ratios are complicated functions of the mixing angles,
each of them can be simply expressed to be proportional to $\tan^{6} \beta$ for
$\tan\beta \gg 1$ in the
SM-like region ($\sin(\alpha-\beta) \sim -1$).
This $\tan^{6} \beta$ dependence is a common feature of the tau-associated
LFV processes with the Higgs-mediated 4-Fermi interactions\footnote{
This also applies to the one-loop induced $\tau \rightarrow \ell_i \gamma$
processes 
whose diagrams have the same $\tan\beta$ dependence.}.  

The experimental upper limit 
on $|\kappa_{3i}|^{2}$ can be obtained  
by using the experimental results given in Table~\ref{Tab:tau-bound} and analytic
expressions for the decay branching ratios in Eqs.~\eqref{eq:tmpi0}-\eqref{eq:tmg}.
For instance, let us consider the bound 
from the $\tau\rightarrow \mu\eta$ results\cite{Sher-tmeta}. 
From Eq.~\eqref{eq:tmeta}
we obtain 
\begin{align}
&|\kappa_{32}|^{2} \leq
 \left( 
 \left|
  \kappa^{\text{max}}_{32}
 \right|^{2}
 \right)_{\tau^{}\rightarrow \mu^{}\eta}
 \equiv
 \frac{128 \pi {\rm Br}(\tau^{} \rightarrow \mu^{} \eta)_{\rm exp}  m_{A}^{4}}
 {9 G_{F}^{2} m_{\tau}^{3} m_{\eta}^{4} F_{\eta}^{2} \tau_{\tau} 
  \left(
   1- \frac{m_{\eta}^{2}}{m_{\tau}^{2}}
  \right)^{2}}
 \frac{\cos^{6}\beta}{\sin^{2}\beta},
\label{eq:bound-from-tmeta}
\end{align}
where ${\rm Br}(\tau^{} \rightarrow \mu^{} \eta)_{\rm exp}$ is the
experimental upper limit on the branching ratio of $\tau^{} \rightarrow
\mu^{} \eta$ in Table~\ref{Tab:tau-bound}.
In particular, for $\tan\beta\gg1$,
the right-hand-side can be expressed by
\begin{align}
 \left( 
 \left|
  \kappa^{\text{max}}_{32}
 \right|^{2}
 \right)_{\tau^{}\rightarrow \mu^{}\eta}
\simeq
2.3 \times 10^{-4} \times 
\left(
 \frac{m_{A}^{}}{350 {\rm [GeV]}}
\right)^{4}
\left(
\frac{30}{\tan\beta}
\right)^{6}.
\label{eq:kappa32sqmax}
\end{align} 
It can be easily seen that the bound 
$(|\kappa_{32}^{\text{max}}|^{2})_{\tau\rightarrow \mu\eta}$ is rapidly relaxed 
in the region with small $\tan\beta$ and large $m_{A}^{}$. 
In a similar way to  Eq.~\eqref{eq:bound-from-tmeta},
the maximal allowed value
$\left( |\kappa_{3i}^{\text{max}}|^{2} \right)_{\text{mode}}$ can be
calculated for each mode.
The combined upper limit ($|\kappa^{\text{max}}_{3i}|^{2}$) is then given by
\begin{align}
\left|\kappa^{\text{max}}_{3i}\right|^{2}
 \equiv
 {\rm MIN}
 \left\{
  \left(
  \left|
   \kappa^{\text{max}}_{3i}
  \right|^{2}
  \right)_{\tau\rightarrow\ell\eta},
  \left(
  \left|
   \kappa^{\text{max}}_{3i}
  \right|^{2}
  \right)_{\tau \rightarrow \ell \mu^{+}\mu^{-}},
  \left(
  \left|
   \kappa^{\text{max}}_{3i}
  \right|^{2}
  \right)_{\tau \rightarrow \ell K^{+}K^{-}},
  \left(
  \left|
   \kappa^{\text{max}}_{3i}
  \right|^{2}
  \right)_{\tau \rightarrow \ell \gamma}, \cdot\cdot\cdot
 \right\}.
\label{eq:kappaSq-max}
\end{align}
As shown below, 
$\tau^{}\rightarrow \ell^{}_i\eta$ and  
$\tau^{}\rightarrow \ell^{}_i\gamma$
give the strongest upper limits on $|\kappa_{3i}|^{2}$
in a wide range of the parameter space. 
In addition, in some parameter regions  
$\tau^{}\rightarrow \ell^{}_iK^{+}K^{-}$
and $\tau^{}\rightarrow \ell^{}_i\mu^{+}\mu^{-}$ can 
also give similar limits on $|\kappa_{3i}|^{2}$ 
to those from the above two processes.

\begin{figure}[t]
\begin{minipage}{8cm}
\unitlength=1cm
\begin{picture}(7.8,6)
\put(0,1){\includegraphics[width=8cm]{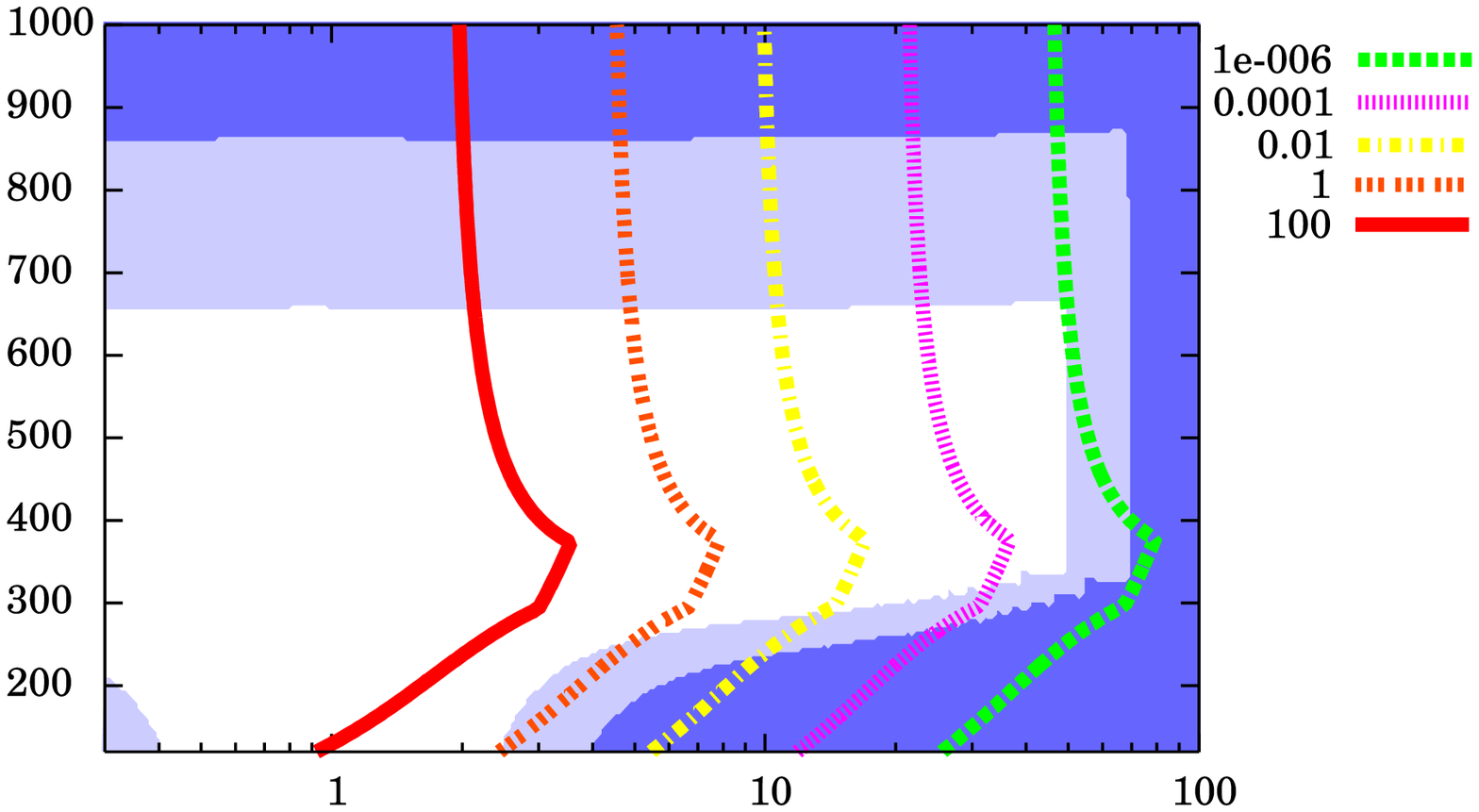}}
\put(0,5.5){$m_{A}^{}$ [GeV]}
\put(6.8,1){$\tan\beta$}
\put(3,0.5){(a)}
\end{picture}
\end{minipage}
\begin{minipage}{8cm}
\unitlength=1cm
\begin{picture}(7.8,6)
\put(0,1){\includegraphics[width=7.8cm]{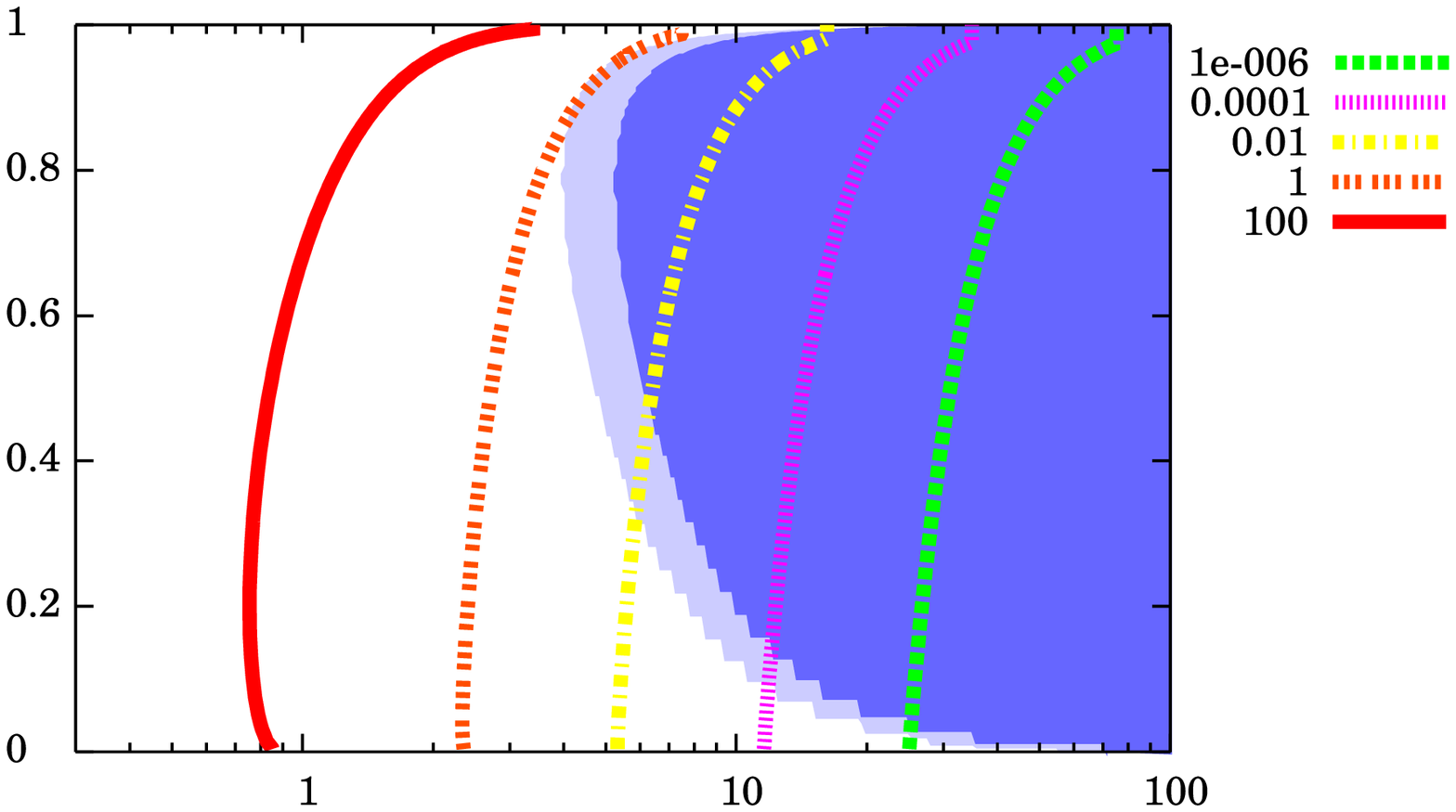}}
\put(0,5.6){$\sin^{2}(\alpha-\beta)$}
\put(7,1){$\tan\beta$}
\put(3,0.5){(b)}
\end{picture}
\end{minipage}
\caption{
Contours of  $|\kappa^{\text{max}}_{32}|^{2}$,  
the possible maximal value of $|\kappa_{32}|^{2}$ from the rare tau
decay results, are shown (a) in the $\tan\beta$-$m_{A}$ plane and (b) 
in the $\tan\beta$-$\sin^{2}(\alpha-\beta)$ plane.
The parameters are taken to be (a) $m_{h}^{} =120$ GeV, 
$m_{H}^{} = m_{H^{\pm}} = 350$ GeV and 
$\sin(\alpha -\beta) = -0.9999$, 
and (b) $m_{h}^{} =120$ GeV and $m_{H}^{} = m_{A} = m_{H^{\pm}} = 350$ GeV.
The remaining parameter $M$ is scanned from 0 to 1,000 GeV. 
The dark (light) shaded area indicates 
the excluded region by the theoretical requirements of 
Eqs.~(\ref{eq:VS}), (\ref{eq:PU}) and (\ref{eq:yukawa-bound}) 
with $\xi=1$ ($\xi=1/2$).
}
\label{Fig:allowed-kappa}
\end{figure}

In Figs.~\ref{Fig:allowed-kappa}-(a) and \ref{Fig:allowed-kappa}-(b),
contour plots for $|\kappa_{32}^{\text{max}}|^{2}$ are shown 
under the rare tau decay results in the $\tan\beta$-$m_{A}$ plane and
the $\tan\beta$-$\sin^{2}(\alpha-\beta)$ plane, respectively. 
The combined excluded region from the theoretical conditions in 
Eqs.~\eqref{eq:VS}, \eqref{eq:PU} and \eqref{eq:yukawa-bound}  
is indicated by the dark shaded area for 
the criterion $\xi=1$ and by the light one for $\xi=1/2$.
In Fig.~\ref{Fig:allowed-kappa}-(a), 
parameters of the Higgs sector are taken to be
$m_{h}^{} =120$ GeV, $m_{H}^{} = m_{H^{\pm}}^{} = 350$ GeV and 
$\sin(\alpha -\beta) = -0.9999$.
In Fig.~\ref{Fig:allowed-kappa}-(b), those are 
$m_{h}^{} =120$ GeV and $m_{H}^{} = m_{A}^{} = m_{H^{\pm}}^{} = 350$ GeV.
The value of $|\kappa^{\text{max}}_{32}|^{2}$ is independent of 
$M$, the soft-breaking scale of the discrete symmetry. 
On the other hand, theoretical bounds from vacuum stability and 
perturbative unitarity are sensitive to $M$.
Therefore, we evaluate such a theoretical allowed region by scanning 
$M$ to be from 0 to 1000 GeV.
We also take into account the constraint from the $\rho$ parameter 
measurement and the $b\rightarrow s \gamma$ result
by taking $\sin(\alpha-\beta) \simeq -1$ and 
$m_{H}^{}=m_{H^{\pm}}^{}$ with $m_{H^{\pm}}^{}\gtrsim 350$ 
GeV for Fig~\ref{Fig:allowed-kappa}-(a), 
and $m_{A}^{}=m_{H^{\pm}}^{}$ with $m_{H^{\pm}}^{}\gtrsim350$ GeV 
for Fig~\ref{Fig:allowed-kappa}-(b). 
From the both figures, 
it is easily found that
the value of $|\kappa_{32}^{\text{max}}|^{2}$
can extensively be larger for smaller $\tan\beta$
in the allowed region under the theoretical constraints.
For $\tan\beta \lesssim 10$ $(30)$,  $|\kappa_{32}^{\text{max}}|^{2}$ 
can be $\mathcal{O}(0.1)$  ($\mathcal{O}(10^{-4})$).
Among the rare tau decay processes, $\tau \rightarrow \mu\eta$ and
$\tau \rightarrow \mu \gamma$ provide the most stringent
constraints on $|\kappa_{32}^{}|^{2}$.
While $\tau \rightarrow \mu \eta$ is mediated only by $A$,
$\tau \rightarrow \mu\gamma$ depends on the masses of $h$, $H$, 
$A$ and $H^\pm$.
For $\sin^{2}(\alpha-\beta) \sim 1$ and $m_{A}^{} \sim m_{H}^{}$, 
the branching ratio of $\tau\to\mu\gamma$ in Eq.~(\ref{eq:tmg}) 
is suppressed because of the cancellation between the one-loop diagrams 
of $A$ and $H$. 
Therefore, $|\kappa_{32}|^2$ is bounded most strongly 
by the $\tau \rightarrow \mu \eta$ result for this 
case\footnote{The MSSM result approximately corresponds to 
this case\cite{Sher-tmeta}.}.
When $m_A$ differs from $m_H$ or when $\sin^2(\alpha-\beta)$ is to some
extent smaller than unity, the one-loop induced 
$\tau\rightarrow \mu\gamma$ process becomes important, 
and gives the most stringent bound on $|\kappa_{32}|^2$ 
of all the tau rare decay processes.

\begin{figure}[t]
\begin{minipage}{8cm}
\unitlength=1cm
\begin{picture}(7.8,6)
\put(0,1){\includegraphics[width=8cm]{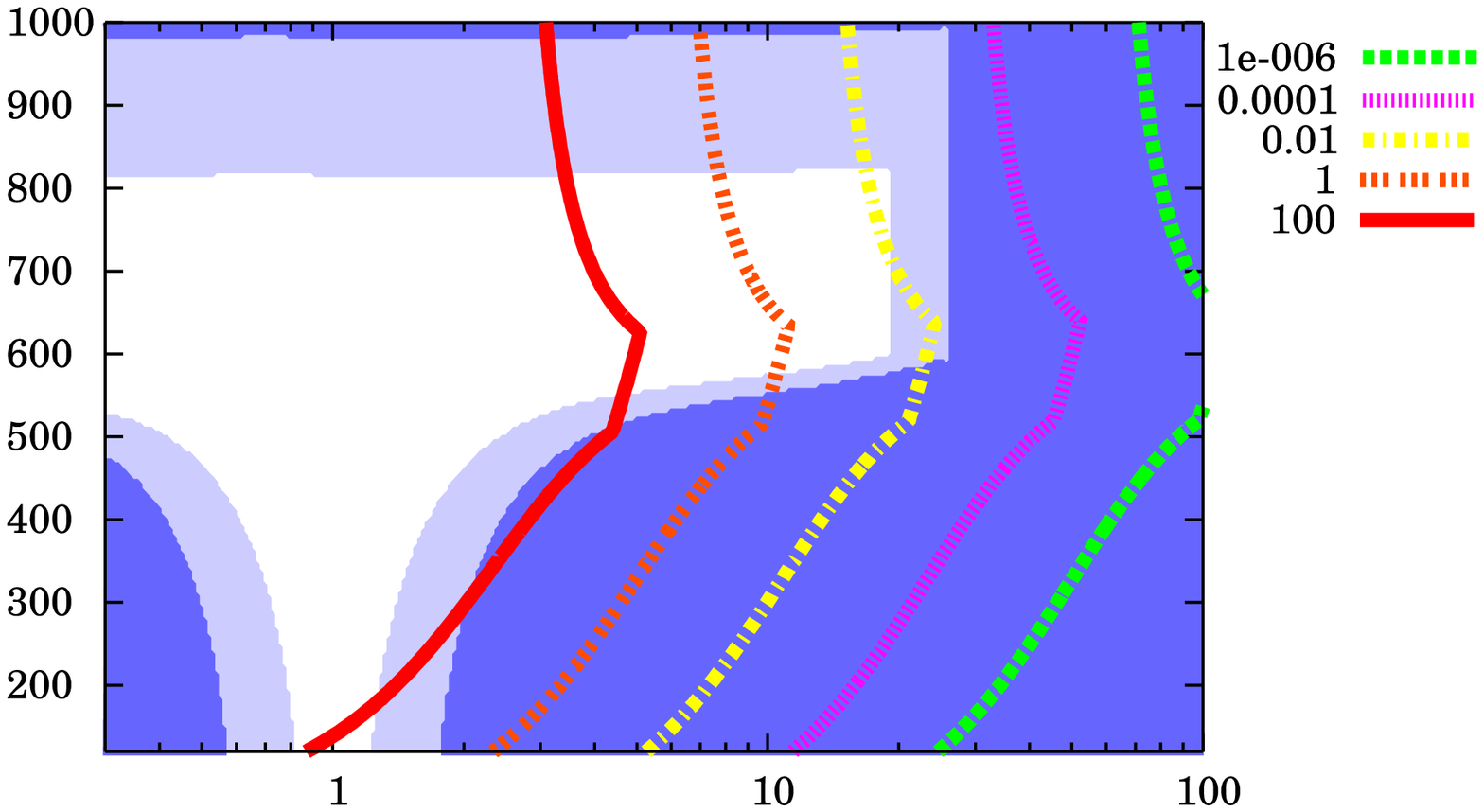}}
\put(0,5.5){$m_{A}^{}$ [GeV]}
\put(6.8,1){$\tan\beta$}
\put(3,0.5){(a)}
\end{picture}
\end{minipage}
\begin{minipage}{8cm}
\unitlength=1cm
\begin{picture}(7.8,6)
\put(0,1){\includegraphics[width=7.8cm]{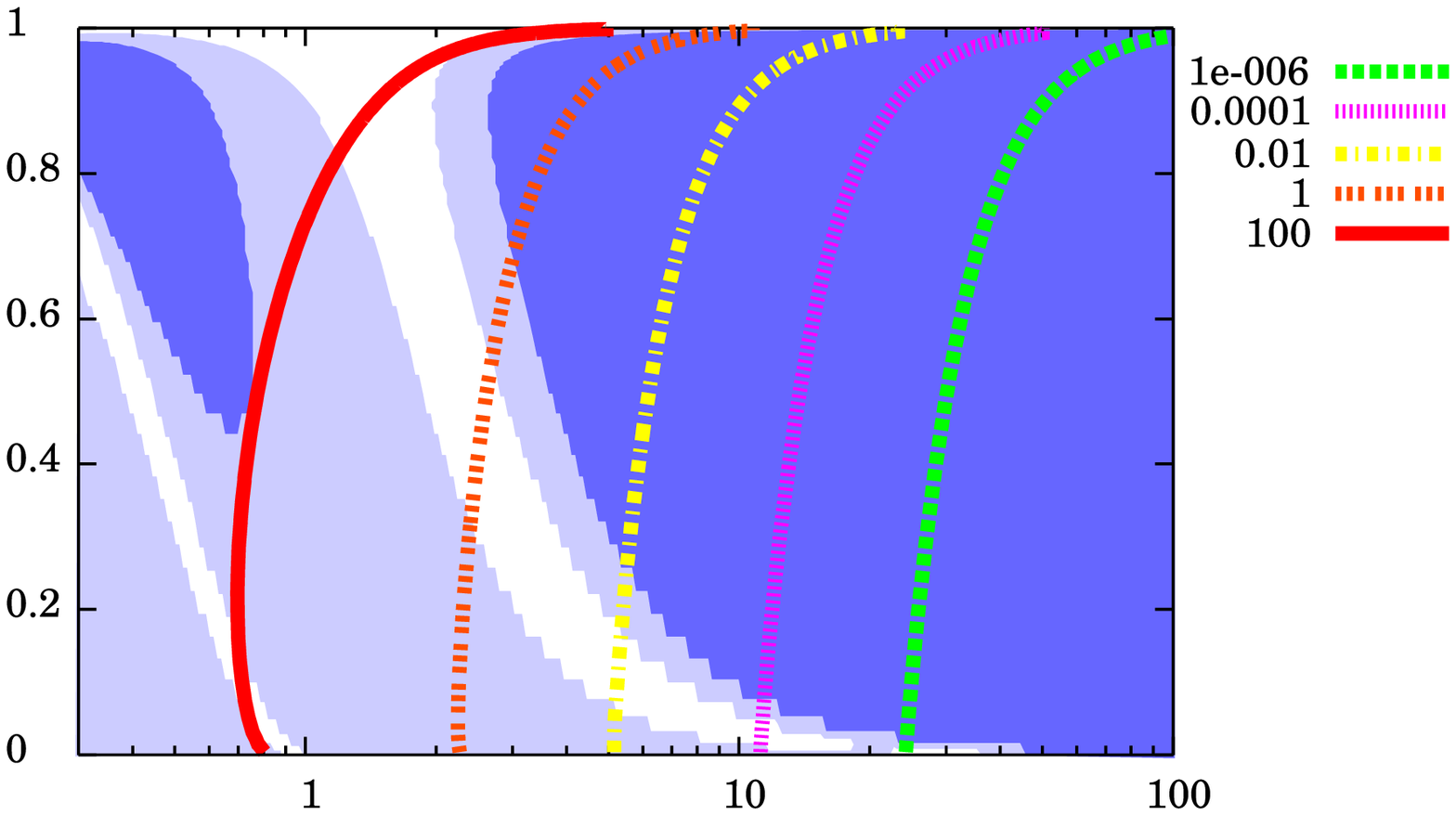}}
\put(0,5.6){$\sin^{2}(\alpha-\beta)$}
\put(7,1){$\tan\beta$}
\put(3,0.5){(b)}
\end{picture}
\end{minipage}
\caption{
Contour plots of $|\kappa^{\text{max}}_{32}|^{2}$ 
are shown (a) in the $\tan\beta$-$m_{A}$ plane and (b) in the
$\tan\beta$-$\sin^{2}(\alpha-\beta)$ plane 
similar to Figs. \ref{Fig:allowed-kappa}-(a) 
and \ref{Fig:allowed-kappa}-(b) but with 
different parameter choices; 
(a) $m_{h}^{} =120$ GeV,
$m_{H}^{} = m_{H^{\pm}} = 600$ GeV and $\sin(\alpha -\beta) = -0.9999$, 
and (b) $m_{h}^{} =120$ GeV and $m_{H}^{} = m_{A} = m_{H^{\pm}} = 600$ GeV.
The dark (light) shaded area indicates 
the excluded region by the theoretical requirements of 
Eqs.~(\ref{eq:VS}), (\ref{eq:PU}) and (\ref{eq:yukawa-bound}) 
with $\xi=1$ ($\xi=1/2$).
}\label{Fig:allowed-kappa-No2}
\end{figure}

In Fig.~\ref{Fig:allowed-kappa-No2}-(a) and 
\ref{Fig:allowed-kappa-No2}-(b), the similar contour plots are shown with
assuming larger values of $m_{H}^{}$ and $m_{{H}^{\pm}}^{}$; i.e.,
$m_{H}^{}=m_{H^{\pm}} = 600$ GeV with $\sin(\alpha-\beta) =-0.9999$
for Fig.~\ref{Fig:allowed-kappa-No2}-(a)
and $m_{H}=m_{A}^{}=m_{H^{\pm}} = 600$ GeV for Fig.~\ref{Fig:allowed-kappa-No2}-(b), 
respectively.
The magnitude of $|\kappa^{\text{max}}_{32}|^{2}$ 
and its $\tan\beta$ dependence are similar
to the case of Fig.~\ref{Fig:allowed-kappa}-(a) 
and \ref{Fig:allowed-kappa}-(b),
although the theoretical allowed regions become changed to the
considerable extent.

In Figs.~\ref{Fig:allowed-kappa} and \ref{Fig:allowed-kappa-No2}, 
the value of $|\kappa^{\text{max}}_{32}|^{2}$ can be
much larger than 100 in a wide range of the parameter region. 
One might think that 
such large values of $|\kappa_{3i}|$ cannot be consistent with the
unitarity argument for the LFV Yukawa couplings.
However, 
it should be emphasized that the above figures show 
the contour plots for $|\kappa_{32}^{\text{max}}|^{2}$ under the
rare tau decay results, and not for $|\kappa_{32}|^{2}$.
The region of $|\kappa^{\text{max}}_{32}|^{2} \gtrsim 1$ should be
taken as the area where $|\kappa_{32}|^{2} $ can be  
as large as $\mathcal{O}(10^{-2}\text{-}10^{-4})$ easily. 
It is concluded that current results of the tau LFV decays do not give 
any substantial upper limit on $|\kappa_{32}|^2$ except for 
high $\tan\beta$ region ($\tan\beta \gtrsim 30$). 

Finally, we comment on the bound on $|\kappa_{31}|^2$, 
the LFV parameters for $\tau$-$e$ mixing.
Similar to $\tau$-$\mu$ mixing, we can discuss 
$|\kappa_{31}^{\rm max}|^2$ 
comparing the data of $\tau \to e \eta$, $\tau \to e \mu^+\mu^-$, 
$\tau \to e K^+K^-$ and $\tau \to e \gamma$ listed 
in Table~\ref{Tab:tau-bound} 
with the formulas given in 
Eqs.~(\ref{eq:tmeta})-(\ref{eq:tmg}).
These formulas for $\tau$-$e$ mixing 
are common with $\tau$-$\mu$ mixing 
except for the factor of $|\kappa_{3i}|^2$, so that 
difference in contours of $|\kappa_{31}^{\rm max}|^2$ from 
those of $|\kappa_{32}^{\rm max}|^2$ only comes from that in the data. 
In Table~\ref{Tab:tau-bound},
the experimental limit for the branching ratio of $\tau \to e \eta$ is 
about 1.5 times weaker than that of $\tau \to \mu \eta$, while  
that of $\tau \to e \gamma$ is 5.7 times relaxed as compared to
that of $\tau \to \mu \gamma$. 
Moreover, the upper limit on ${\rm Br}(\tau^- \to e^-K^+K^-)$ is 
1.8 times stronger than that on ${\rm Br}(\tau^- \to \mu^-K^+K^-)$.
We have numerically confirmed that 
there are some regions where $\tau^- \to e^-K^+K^-$ can give 
the most stringent bound on $|\kappa_{31}|^2$. 
Therefore, $|\kappa_{31}^{\rm max}|^2$ is determined from one of 
$\tau \to e \eta$, $\tau \to e \gamma$ and $\tau^- \to e^-K^+K^-$ 
depending on parameter regions.

\section{Lepton flavor violating Higgs boson decays}
\label{Sec:LFV-Hdecay}

As shown in the previous section,
the LFV Yukawa couplings can be tested 
only in the large $\tan\beta$ region by searching for rare tau decays. 
In order to cover the region unconstrained by rare tau decay results, 
we here consider LFV via the decay of the neutral Higgs bosons;
i.e., $\phi^{0} \rightarrow \tau^\pm \ell_i^\mp$ ($\phi^{0}=h,H$ and $A$).
Branching ratios for these decays are 
calculated\cite{Rossi,Assamagan,Osaka,Herrero} to be
\begin{align}
{\rm Br}(h\rightarrow \tau^{-} \ell^{+}_i) =&
 \frac{1}{16\pi} 
 \frac{m_{\tau}^{2} \cos^{2} \left(\alpha - \beta \right)}
 {v^{2} \cos^{4}\beta}
 \left| \kappa_{3i} \right|^{2}
 \frac{m_{h}^{} \left(1- \frac{m_{\tau}^{2}}{m_{h}^{2}}\right)^{2}}
 {\Gamma(h\rightarrow\text{all})}, 
\label{eq:Brhtm}\\
{\rm Br}(H\rightarrow \tau^{-} \ell^{+}_i) =&
 \frac{1}{16\pi} 
 \frac{m_{\tau}^{2} \sin^{2} \left(\alpha - \beta \right)}
 {v^{2} \cos^{4}\beta}
 \left| \kappa_{3i} \right|^{2}
 \frac{m_{H}^{} \left(1- \frac{m_{\tau}^{2}}{m_{H}^{2}}\right)^{2}}
 {\Gamma(H\rightarrow\text{all})},
\label{eq:BrhtmH}\\
{\rm Br}(A\rightarrow \tau^{-} \ell^{+}_i) =&
 \frac{1}{16\pi} 
 \frac{m_{\tau}^{2}}
 {v^{2} \cos^{4}\beta}
 \left| \kappa_{3i} \right|^{2}
 \frac{m_{A}^{} \left(1- \frac{m_{\tau}^{2}}{m_{A}^{2}}\right)^{2}}
 {\Gamma(A\rightarrow\text{all})},
\label{eq:BrhtmA}\end{align}
where $\Gamma (\phi^{0} \rightarrow \text{all}) $  
is the total width for corresponding neutral Higgs boson $\phi^{0}$.
We here neglect terms of $\mathcal{O}(m_{\ell_i}^{2}/m_{\phi^{0}}^{2})$.
Branching ratios for $\phi^0 \to \tau^+\ell_i^-$ coincide with 
those for $\phi^0 \to \tau^-\ell_i^+$ given in 
Eqs.~(\ref{eq:Brhtm}), (\ref{eq:BrhtmH}) and (\ref{eq:BrhtmA}).
In the following, we concentrate on the decays into a $\tau$-$\mu$
pair. We take the values of the SM parameters as 
$\alpha_{\text{em}}=0.007297$, $G_{F}=1.166 \times 10^{-5}$
$\text{GeV}^{-2}$,
$m_{Z}^{}=91.19$ GeV,
$m_{\tau} = 1.777$ GeV, $m_{\mu}=0.1057$ GeV, 
$m_{b}=4.1$ GeV, $m_{t}=174.3$ GeV, and $m_{c}=1.15$ GeV.

\subsection{The LFV decay of the lightest Higgs boson}

A search for the LFV decays $h\rightarrow \tau^\pm\ell_i^\mp$ 
can give important information for extended Higgs sectors and thus 
for the structure of new physics,
even when only $h$ is found and any other direct signals for the extended
Higgs sector are not obtained by experiments.
We here evaluate the possible maximal value 
of the branching ratio 
${\rm Br}(h \rightarrow \tau^{-} \mu^{+})_{\text{max}}$
under the results of the rare tau decay search,
by inserting $|\kappa_{32}^{\text{max}}|^{2}$ 
of Eq.~\eqref{eq:kappaSq-max} into
the $|\kappa_{32}|^{2}$ in Eq.~\eqref{eq:Brhtm}. 
In calculation of $\Gamma(h\to {\rm all})$, the decay modes 
of $b \bar b$, $c \bar c$, $\tau^+ \tau^-$, $\gamma\gamma$, 
$gg$, $Z\gamma$\cite{HHG} and $\tau^\pm \ell_i^\mp$ are taken into 
account. 

\begin{figure}[t]
\begin{minipage}{8cm}
\unitlength=1cm
\begin{picture}(7.8,6)
\put(0,1){\includegraphics[width=8cm]{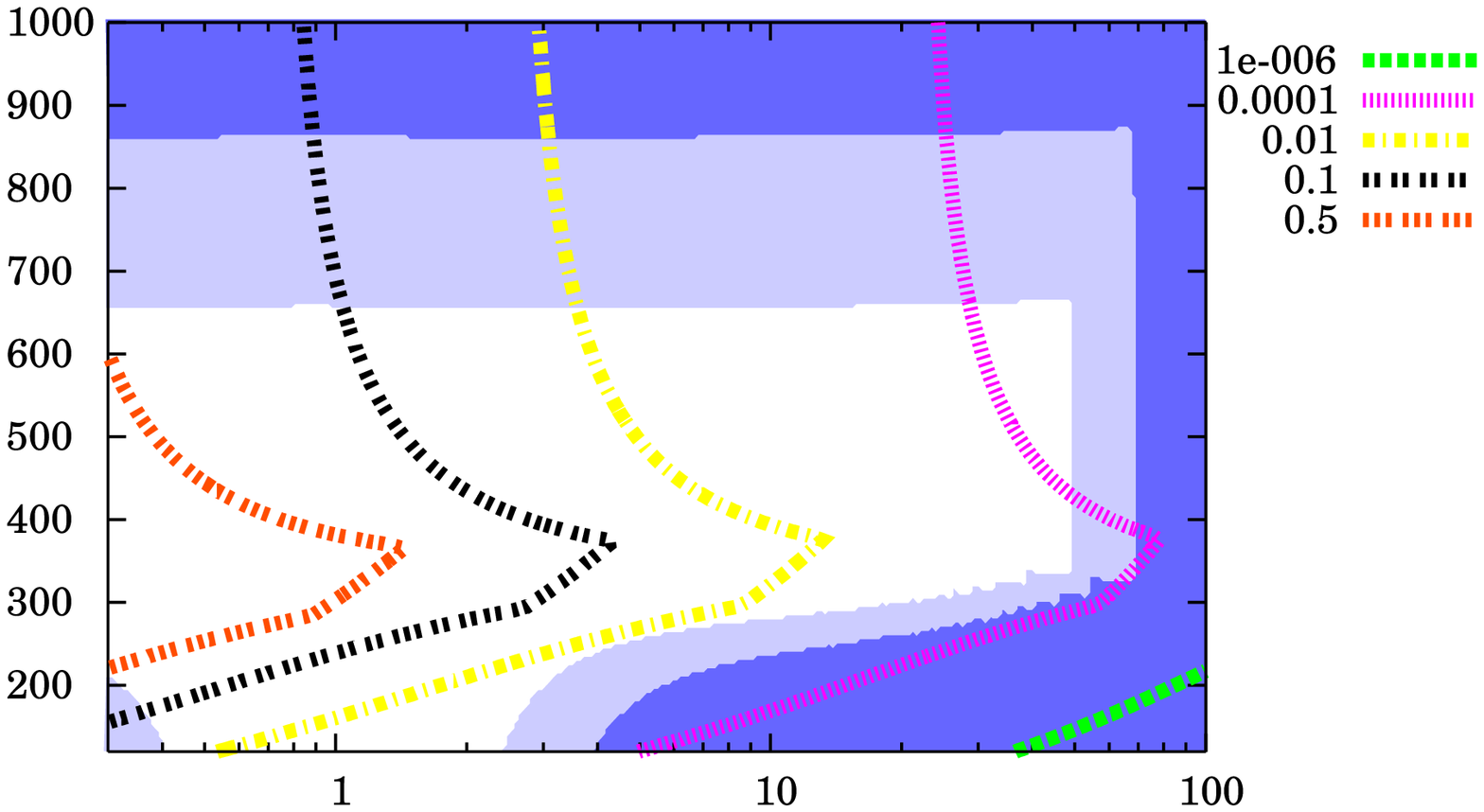}}
\put(3,0.5){(a)}
\put(0,5.5){$m_{A}^{}$ [GeV]}
\put(6.8,1){$\tan\beta$}
\end{picture}
\end{minipage}
\begin{minipage}{8cm}
\unitlength=1cm
\begin{picture}(7.8,6)
\put(0,1){\includegraphics[width=7.8cm]{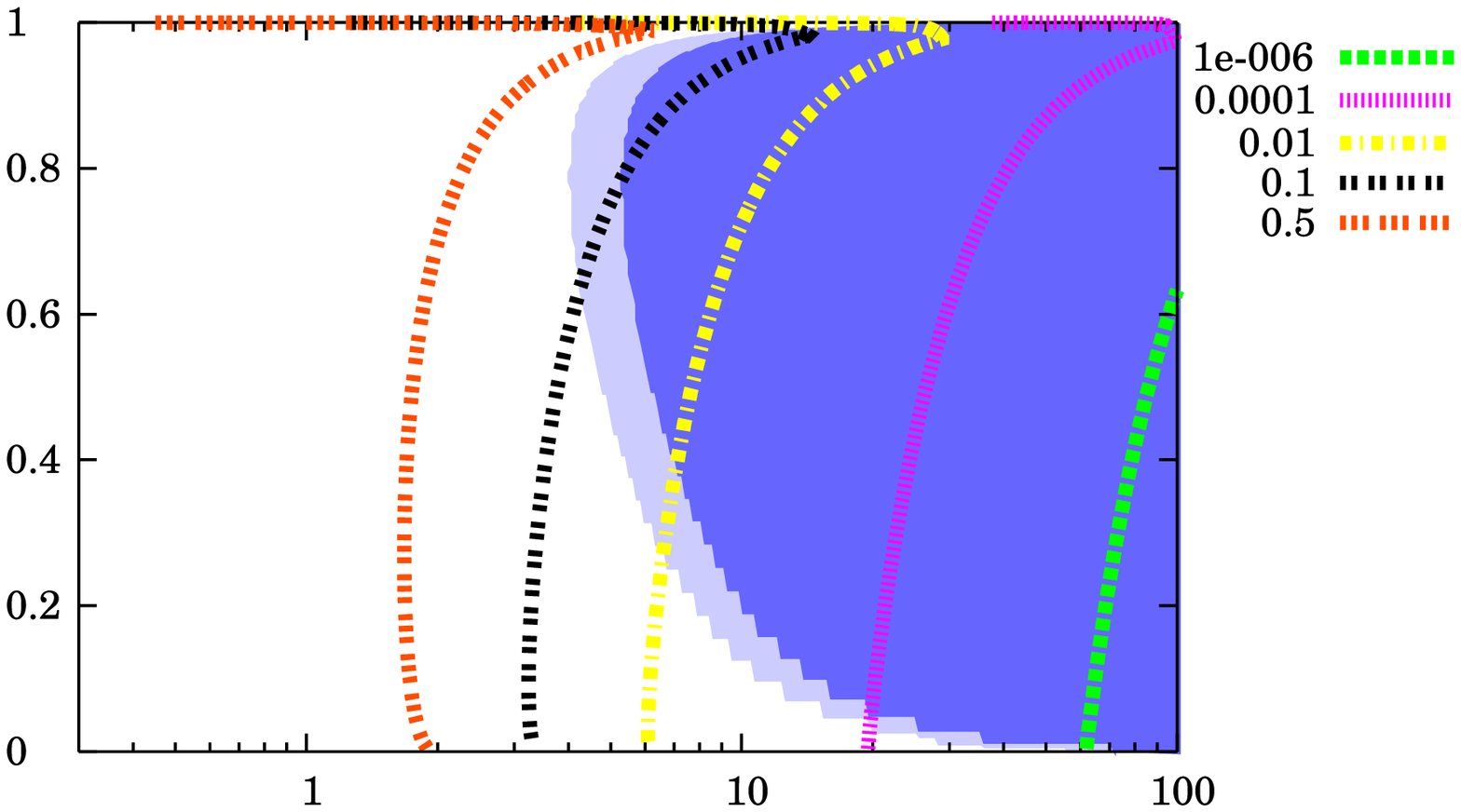}}
\put(3,0.5){(b)}
\put(0,5.5){$\sin^{2}(\alpha-\beta)$}
\put(6.8,1){$\tan\beta$}
\end{picture}
\end{minipage}
\caption{Contour plots of 
 ${\rm Br}(h \rightarrow \tau^{\pm} \mu^{\mp})_{\text{max}}$, 
 the possible maximal values for the branching ratio 
 under the tau rare decay results, are shown (a) in  
 the $\tan\beta$-$m_{A}$ plane  
 and (b) in the $\tan\beta$-$\sin(\alpha-\beta)$ plane. 
 The parameters are taken as the same as 
 Figs.~\ref{Fig:allowed-kappa}-(a) and \ref{Fig:allowed-kappa}-(b), 
 respectively. 
 The dark (light) shaded area indicates 
 the excluded region by the theoretical requirements of 
Eqs.~(\ref{eq:VS}), (\ref{eq:PU}) and (\ref{eq:yukawa-bound}) with $\xi=1$ 
($\xi=1/2$).
}
\label{Fig:Br-htm}
\end{figure}

In Figs.~\ref{Fig:Br-htm}-(a) and \ref{Fig:Br-htm}-(b), contours of 
${\rm Br}(h \rightarrow \tau^{\pm}\mu^{\mp})_{\text{max}}$, which is 
twice of ${\rm Br}(h \rightarrow \tau^{-}\mu^{+})_{\text{max}}$, 
are shown in the $\tan\beta$-$m_{A}^{}$ plane and
in the $\tan\beta$-$\sin^{2}(\alpha-\beta)$ plane, respectively.
The parameters are taken to be the same as those for
Figs.~\ref{Fig:allowed-kappa}-(a) and \ref{Fig:allowed-kappa}-(b), 
respectively; i.e.,  
(a) $m_{h}^{} =120$ GeV, $m_{H}^{} = m_{H^{\pm}}^{} = 350$ GeV and 
    $\sin(\alpha -\beta) = -0.9999$, and 
(b) $m_{h}^{} =120$ GeV and $m_{H}^{} = m_{A}^{} = m_{H^{\pm}}^{} = 350$
GeV, with $M$ to be scanned from 0 to 1000 GeV.
We again show the excluded area from requirements of tree-level
unitarity and vacuum stability as in the same way 
as Figs.~\ref{Fig:allowed-kappa}-(a) and \ref{Fig:allowed-kappa}-(b). 
For low and moderate values of $\tan\beta$  ($\tan\beta \lesssim 30$),  
where rare tau decay results cannot give
substantial upper limit on $|\kappa_{32}|^{2}$,
${\rm Br}(h \rightarrow \tau^{\pm}\mu^{\mp})_{\text{max}}$ 
can be sufficiently large. 
We find that the possible maximal values of the 
branching ratio can be greater than $\mathcal{O}(10^{-3})$ 
in a wide rage of the allowed region under the theoretical conditions in 
Eqs.~(\ref{eq:VS}), (\ref{eq:PU}) and (\ref{eq:yukawa-bound}). 

\begin{figure}[t]
\begin{minipage}{8cm}
\unitlength=1cm
\begin{picture}(7.8,6)
\put(0,1){\includegraphics[width=8cm]{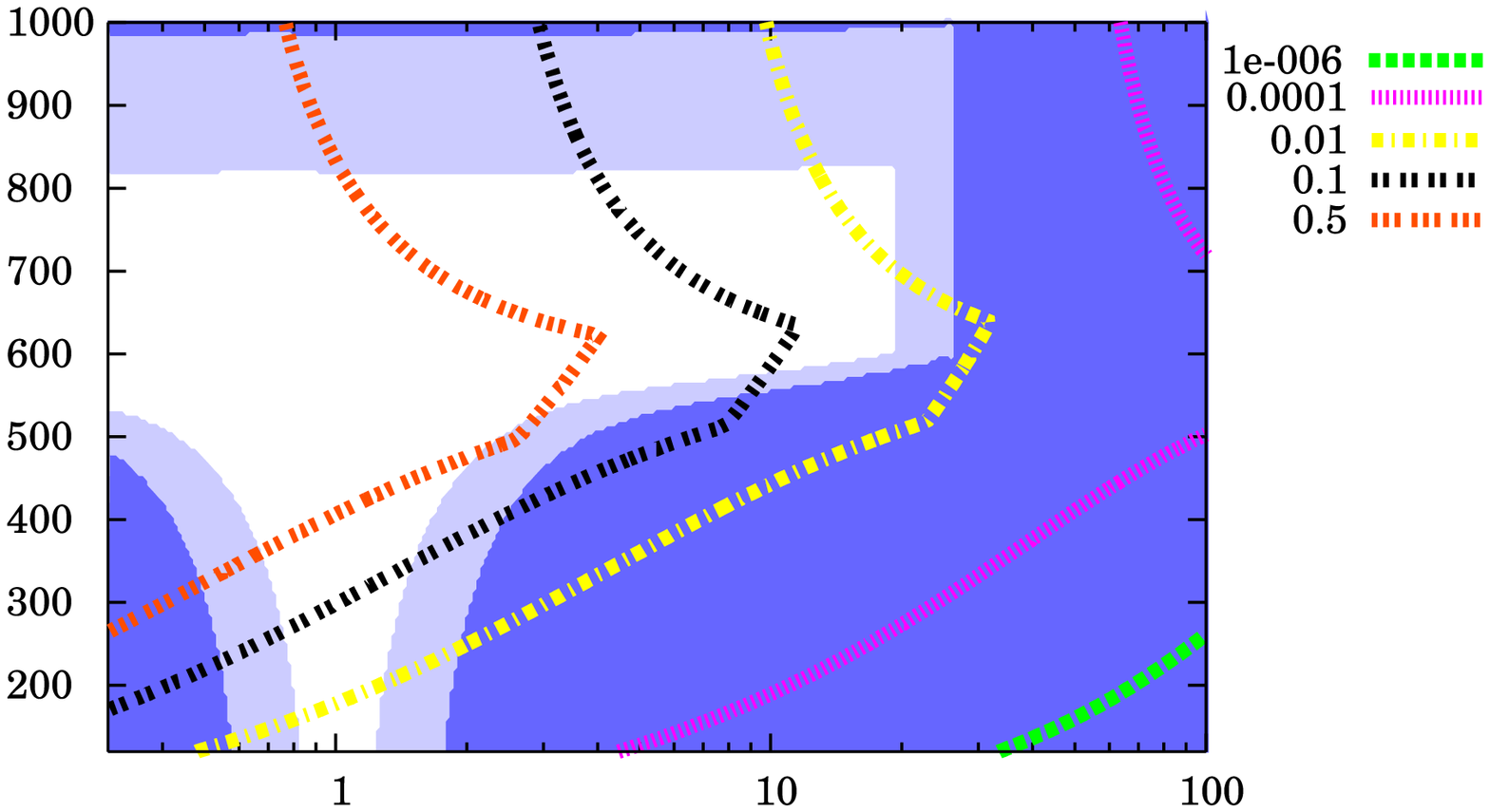}}
\put(3,0.5){(a)}
\put(0,5.5){$m_{A}^{}$ [GeV]}
\put(6.8,1){$\tan\beta$}
\end{picture}
\end{minipage}
\begin{minipage}{8cm}
\unitlength=1cm
\begin{picture}(7.8,6)
\put(0,1){\includegraphics[width=7.8cm]{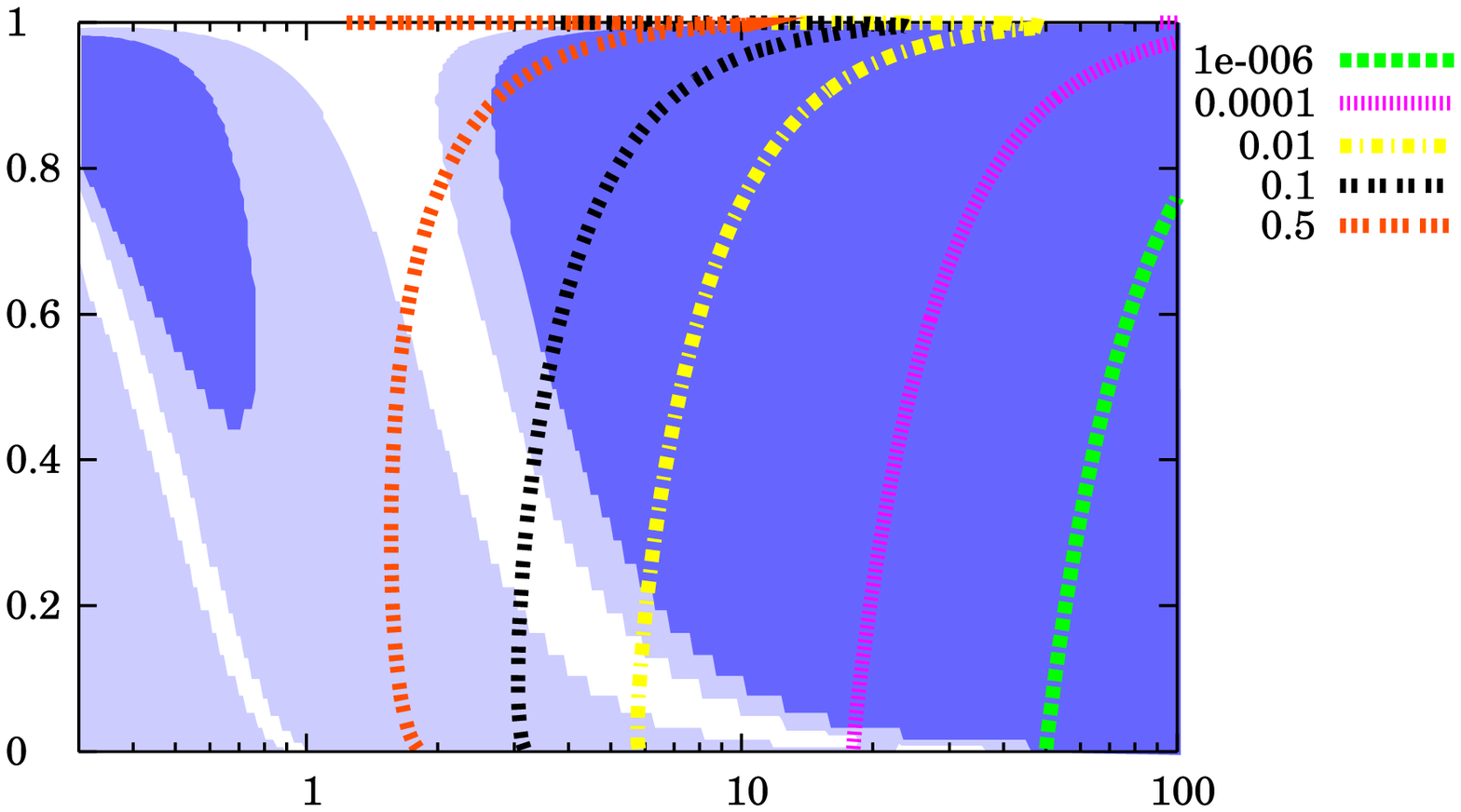}}
\put(3,0.5){(b)}
\put(0,5.5){$\sin^{2}(\alpha-\beta)$}
\put(6.8,1){$\tan\beta$}
\end{picture}
\end{minipage}
\caption{
Contour plots 
of  ${\rm Br}(h \rightarrow \tau^{\pm} \mu^{\mp})_{\text{max}}$ 
similar to Figs.~\ref{Fig:Br-htm}-(a) and \ref{Fig:Br-htm}-(b) 
are shown (a) in the $\tan\beta$-$m_{A}$ plane and (b) 
in the $\tan\beta$-$\sin^{2}(\alpha-\beta)$ plane 
with the parameters taken as the same as 
Figs.~\ref{Fig:allowed-kappa-No2}-(a) and
 \ref{Fig:allowed-kappa-No2}-(b), respectively.
 The dark (light) shaded area indicates 
 the excluded region by the theoretical requirements of 
Eqs.~(\ref{eq:VS}), (\ref{eq:PU}) and (\ref{eq:yukawa-bound}) with $\xi=1$ 
($\xi=1/2$).
}
\label{Fig:Br-htm-No2}
\end{figure}

In Fig.~\ref{Fig:Br-htm-No2}-(a) and 
\ref{Fig:Br-htm-No2}-(b), similar contour plots of 
${\rm Br}(h \rightarrow \tau^{\pm} \mu^{\mp})_{\text{max}}$ 
are shown for the same parameters as 
Figs.~\ref{Fig:allowed-kappa-No2}-(a) and
\ref{Fig:allowed-kappa-No2}-(b), respectively; i.e., 
(a) $m_{H}^{}=m_{H^{\pm}} = 600$ GeV with $\sin(\alpha-\beta) =-0.9999$, 
and (b) $m_{H}=m_{A}^{}=m_{H^{\pm}} = 600$ GeV, with $M$ to be scanned 
from 0 to 1000 GeV.
The excluded area from requirements of tree-level
unitarity and vacuum stability are also shown.
The possible maximal values for the branching ratio 
are similar to those in the case shown in 
Figs.~\ref{Fig:Br-htm}-(a) and \ref{Fig:Br-htm}-(b). 
In the allowed region under the conditions 
in Eqs.~(\ref{eq:VS}), (\ref{eq:PU}) and (\ref{eq:yukawa-bound}), 
${\rm Br}(h \rightarrow \tau^{\pm} \mu^{\mp})_{\text{max}}$ 
can be greater than $\mathcal{O}(10^{-2})$. 

For relatively lower $\tan\beta$ values,
the experimental upper limits on $|\kappa_{32}|^2$ 
from rare tau decays are weaker, and 
$\text{Br}(h\rightarrow\tau^{\pm} \mu^{\mp})$ 
can be sufficiently large 
($\gtrsim \mathcal{O}(10^{-3})$ for $m_h \sim 120$ GeV).      
It is expected that a sufficient number of 
such light $h$ can be produced at future 
colliders such as CERN LHC,  
currently planned International Linear Collider (ILC) and 
CERN CLIC. 
It has been pointed out that the decay process $h \to \tau^\pm\mu^\mp$
can easily be detected at ILC with the luminosity of 1 ab$^{-1}$,  
when $m_h \sim 120$ GeV and 
${\rm Br}(h\rightarrow\tau^\pm\mu^\mp) \gtrsim \mathcal{O}(10^{-3})$\footnote{ 
Below that value, the signal would be seriously suffered from 
the backgrounds. In particular, events from 
$h\to\mu^+\mu^-$ and $h\to\tau^+\tau^-$
would be difficult to be separated from the signal.
The branching ratio of $h \to \mu^+\mu^-$ is approximately 
proportional to $m_\mu^2/(N_c m_b^2)$ for a relatively light $h$,
with $N_c$ being the color factor; i.e.,  
${\rm Br}(h\to \mu^+\mu^-) \sim 3 \times 10^{-4}$.
For a large $\tan\beta$ value with $\sin^2(\alpha-\beta) \ll 1$, 
the maximal value of  ${\rm Br}(h \to \tau^\pm \mu^\mp)$
becomes smaller and can be comparable to ${\rm Br}(h \to \mu^+\mu^-)$. 
A simulation study is necessary to clarify this point.}
via the Higgsstrahlung process by using the recoil 
momentum of $Z$ boson\cite{Osaka}. 
Therefore, the LFV search via the decay $h \to \tau^\pm\mu^\mp$ 
at ILC can be complementary to that via rare tau decays 
at (super) B factories, and the both 
cover a wide region of the parameter space 
of the lepton flavor violating THDM.

\subsection{The LFV decay of Heavier Higgs bosons}

Next we discuss branching ratios for the LFV decays of 
heavier Higgs bosons, $H/A \to \tau^\pm \ell_i^\mp$, 
using Eqs.~(\ref{eq:BrhtmH}) and (\ref{eq:BrhtmA}) under the 
current data of LFV rare tau decays. 
In the THDM, there are many possible decay 
modes for $H$ depending on the mass spectrum; i.e.,      
$hh$, $AA$, $hA$, $h\gamma$, $hZ$, $A\gamma$, $AZ$,  
$H^+H^-$ and  $H^\pm W^\mp$ as well as 
$f\bar f$ ($f=t,b,c,\tau$), 
$W^\pm W^\mp$, $ZZ$, $Z\gamma$, $\gamma\gamma$ and $gg$.
The last three modes as well as 
$hA$, $h\gamma$, $A\gamma$ and $hZ$ 
appear through the one-loop induced couplings\footnote{
In the numerical analysis, we included contributions from 
these $Z\gamma$, $\gamma\gamma$ and $gg$ modes in addition to 
all the tree level modes, but neglected the other loop-induced 
modes.}. 
Those for $A$ are 
$hZ$, $HZ$, $H^+H^-$, $H^\pm W^\mp$ and $f\bar f$
at the tree level 
as well as $hh$, $hH$, $HH$, $h\gamma$, $H\gamma$, 
$W^\pm W^\mp$, $ZZ$, $Z\gamma$, $\gamma\gamma$ and $gg$ 
at the one loop level\cite{HHG}. 

The branching ratios for $H/A \to \tau^\pm \ell_i^\mp$ 
are sensitive to the masses of all the Higgs bosons.
Here we consider the case of 
$\sin(\alpha-\beta)=-1$ and  
$m_H^{}=m_A^{}=m_{H^\pm}^{}$ $(\equiv m_\Phi^{})$. 
As discussed in Sec. II, 
the $\rho$ parameter constraint is satisfied 
for this choice. From the $b \to s \gamma$
results, $m_\Phi^{}$ is taken to be greater than $350$ GeV.  
As also discussed in Sec. II, $M$ 
determines the decoupling property of heavier Higgs bosons. 
Although the branching ratios ${\rm Br}(H/A \to \tau^\pm \ell_i^\mp)$ 
are insensitive to $M$ in the present parameter set, its value 
strongly affects the allowed parameter region under the theoretical 
conditions of Eqs.~(\ref{eq:VS}) and (\ref{eq:PU}).
Notice that couplings of $H$ are similar to those of $A$ for 
$\sin(\alpha-\beta)=-1$ where there are no $HVV$ couplings. 
Hence we show the results only for the LFV decays of $H$ below. 
In a general case, the branching ratio of $H \to \tau^\pm \mu^\mp$ 
tends to be smaller than 
that of $A \to \tau^\pm \mu^\mp$ due to the contribution from 
the modes $H \to VV$.

\begin{figure}[t]
\begin{minipage}{5cm}
\unitlength=1cm
\begin{picture}(7.8,7)
\put(0,1){\includegraphics[width=5.14cm]{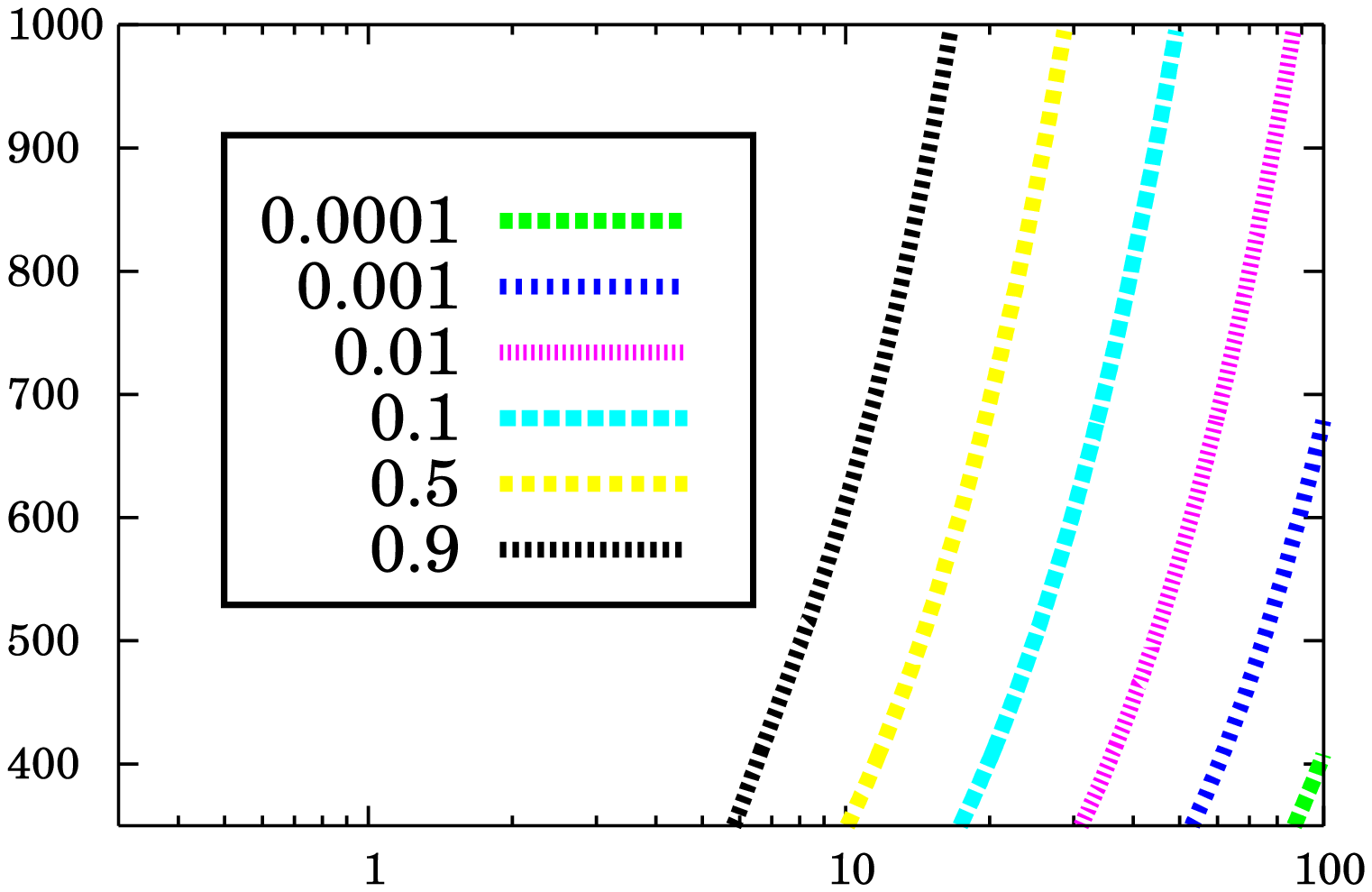}}
\put(2,0.5){(a)}
\put(0,4.6){$m_{\Phi}^{}$ [GeV]}
\put(4,0.7){$\tan\beta$}
\end{picture}
\end{minipage}
\begin{minipage}{5cm}
\unitlength=1cm
\begin{picture}(7.8,7)
\put(0.24,1){\includegraphics[width=4.72cm]{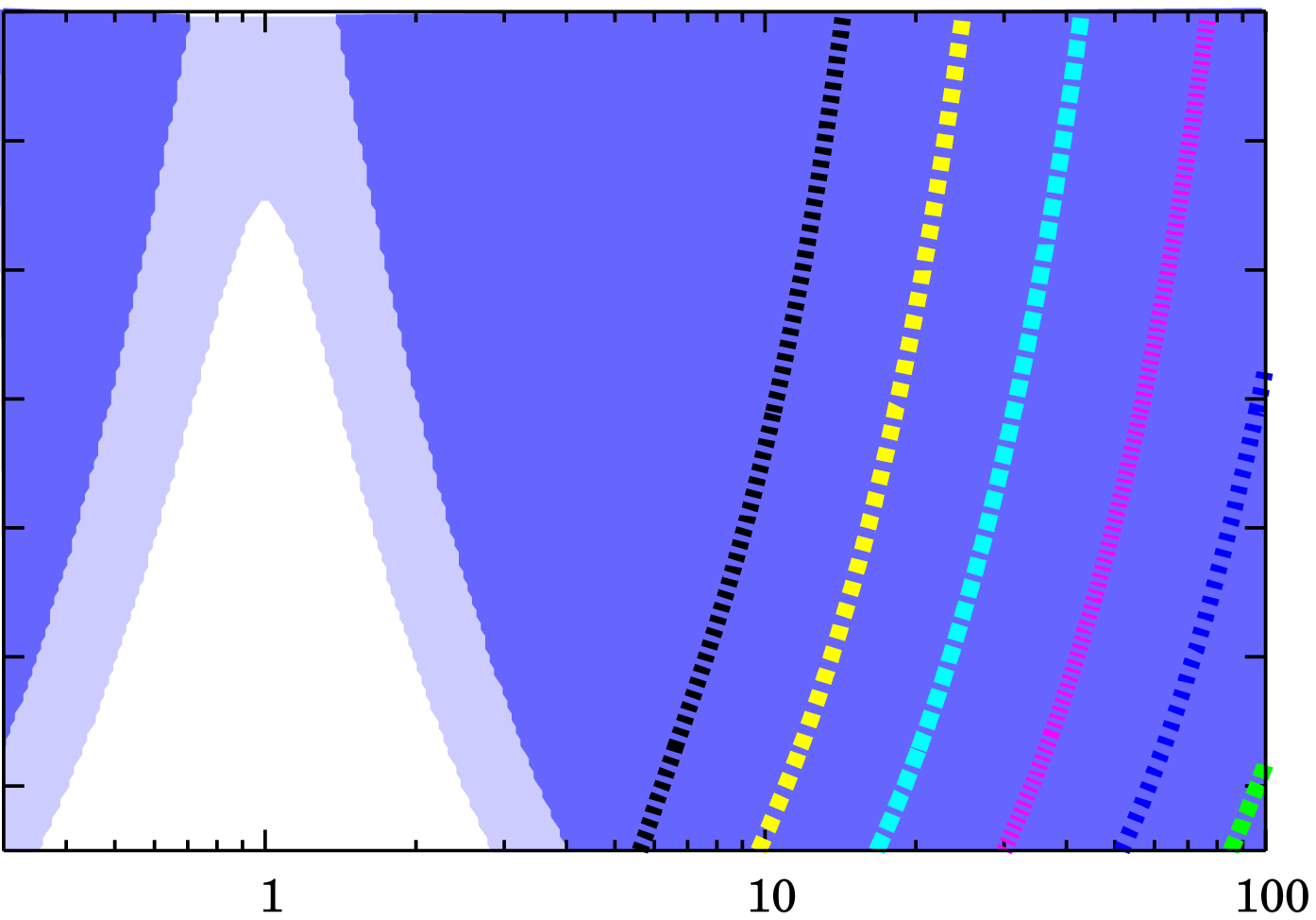}}
\put(2,0.5){(b)}
\put(4,0.7){$\tan\beta$}
\end{picture}
\end{minipage}
\begin{minipage}{5cm}
\unitlength=1cm
\begin{picture}(7.8,7)
\put(0,1){\includegraphics[width=4.71cm]{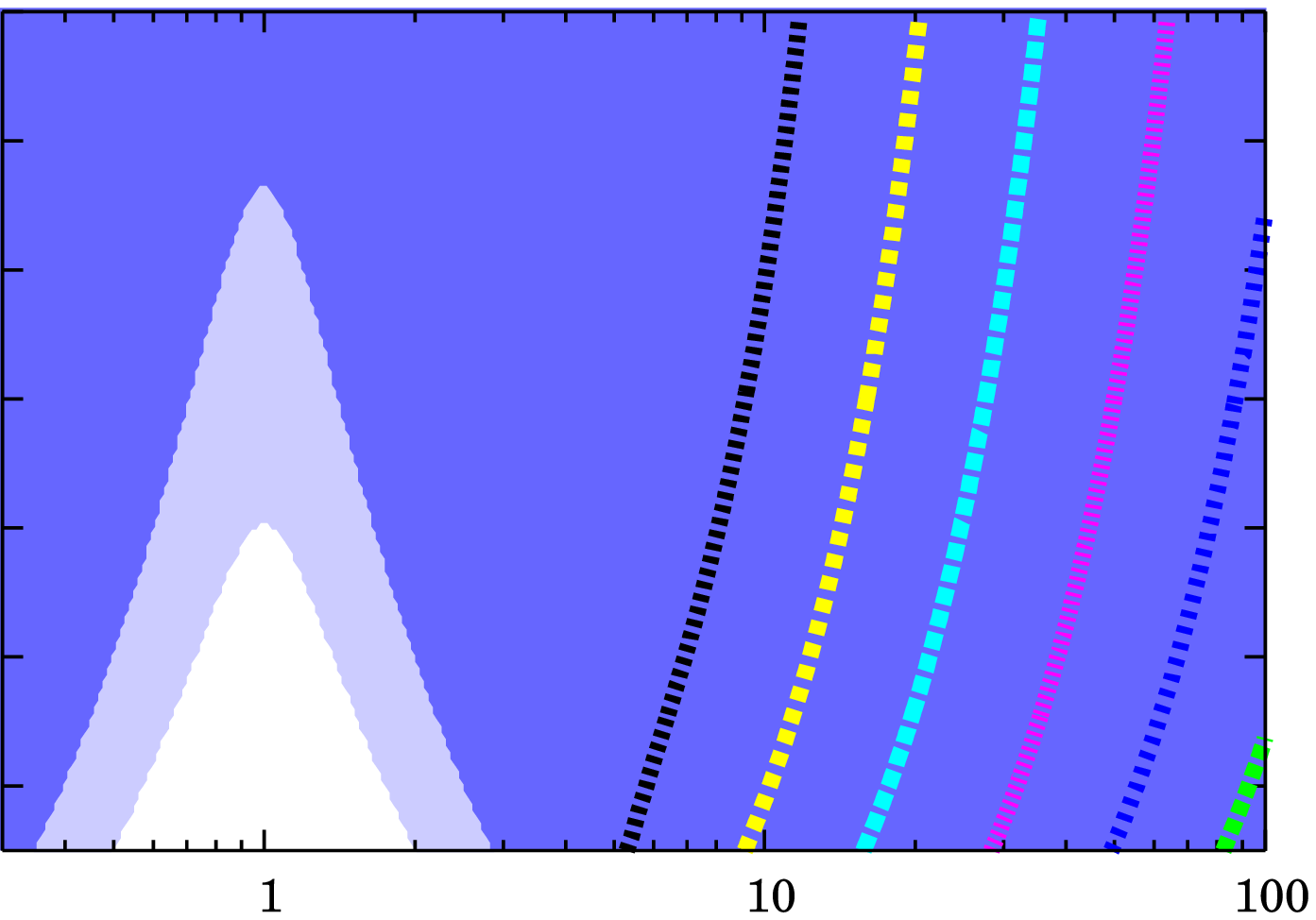}}
\put(2,0.5){(c)}
\put(4,0.7){$\tan\beta$}
\end{picture}
\end{minipage}
\caption{
Contour plots of 
${\rm Br}(H \rightarrow \tau^{\pm} \mu^{\mp})_{\text{max}}$ 
are shown in the $\tan\beta$-$m_{\Phi}^{}$ plane 
($m_\Phi^{}\equiv m_H^{}=m_A^{}=m_{H^\pm}^{}$)
for $m_h^{}=120$ GeV and $\sin(\alpha-\beta)=-1$ with 
(a) $M=m_\Phi^{}$,  
(b) $M=m_\Phi^{}/\sqrt{2}$, and  
(c) $M=0$. 
The dark (light) shaded area indicates 
the excluded region by the theoretical requirements of
Eqs.~(\ref{eq:VS}), (\ref{eq:PU}) and (\ref{eq:yukawa-bound}) 
with $\xi=1$ ($\xi=1/2$).
}
\label{Fig:Br-Htm}
\end{figure}

In Figs.~\ref{Fig:Br-Htm}-(a), 
\ref{Fig:Br-Htm}-(b) and \ref{Fig:Br-Htm}-(c), 
contour plots of ${\rm Br}(H \to \tau^\pm\mu^\mp)_{\rm max}$, 
the upper limit of ${\rm Br}(H \to \tau^\pm\mu^\mp)$ under the rare 
tau decay results, are shown in the $\tan\beta$-$m_\Phi^{}$ 
plane for $M=m_\Phi^{}$, $m_\Phi^{}/\sqrt{2}$ and $0$, respectively.
As expected, the contours are insensitive to the values of 
$M$, and approximately the same in 
Figs.~\ref{Fig:Br-Htm}-(a), \ref{Fig:Br-Htm}-(b) and \ref{Fig:Br-Htm}-(c). 
It is shown that ${\rm Br}(H \to \tau^\pm\mu^\mp)_{\rm max}$ 
can be larger than $10^{-3}$ except for large 
$\tan\beta$ values with relatively small $m_\Phi^{}$. 
Therefore, it turns out to be no substantial upper 
limit on the ${\rm Br}(H \to \tau^\pm\mu^\mp)$ in the 
relatively low $\tan\beta$ region ($\tan\beta \lesssim 20$)
from the LFV rare tau decay results. 
When $M$ is smaller than $m_\Phi$, where 
the heavier Higgs boson partially receive their masses from 
the vacuum expectation value, the allowed parameter region 
is strongly constrained by the requirements 
of vacuum stability and perturbative unitarity.
In particular, for $M=0$ (Fig.~\ref{Fig:Br-Htm}-(c)), 
the allowed region is limited only the area 
of around $\tan\beta \sim 1$ and $m_\Phi^{} \lesssim 600$ GeV. 

The extra Higgs bosons ($H$, $A$ and $H^\pm$) 
are expected to be searched at the LHC. 
The signal of $gg \to H/A \to \tau^\pm \mu^\mp$ 
may be detectable at LHC with high luminosity (100 ${\rm fb}^{-1}$) 
when ${\rm Br}(H/A \to \tau^\pm\mu^\mp)$ is 
greater than $10^{-2}$ for $m_{H/A}^{} \sim 350$ GeV and 
$\tan\beta=45$\cite{Assamagan}. 
However the rate is rapidly reduced for smaller values of $\tan\beta$ 
and for larger values of $m_{H/A}^{}$. 
Further feasibility study is necessary.

\section{Conclusions}
\label{Sec:conclusion}

Lepton flavor violating decays of Higgs bosons 
have been studied in the framework of the THDM, 
in which LFV couplings are introduced as a deviation from 
Model II Yukawa interaction in the lepton sector.
The model parameters are constrained by 
requirements of tree-level unitarity and vacuum stability, 
and also from the experimental results. 
The parameters $|\kappa_{3i}|^2$ in LFV Yukawa interactions  
are bounded from above by using the current data for rare tau LFV decays. 
Each process is mediated by the different combination of the Higgs
bosons, so that the data for each of them provides independent
information to the lepton flavor violating THDM.
It has been found that among the rare tau decay data 
those for 
$\tau^{}\rightarrow \ell^{}_i\eta$ and 
$\tau^{}\rightarrow \ell^{}_i\gamma$
give the most stringent upper limits on $|\kappa_{3i}|^{2}$
in a wide range of the parameter space. 
 
In the large $\tan\beta$ region ($\tan\beta \gtrsim 30 $),  
the upper limit  on $|\kappa_{3i}|^2$ due to the rare tau 
decay data turns out to be 
substantial and comparable with the value predicted 
by assuming some fundamental theories such as SUSY. 
The upper limit would be improved in future by about one order of magnitude 
at the experiment at (super) B factories.
For smaller values of $\tan\beta$, the upper limit
is rapidly relaxed, and no more substantial constraint is obtained from 
the rare tau decay results. 

We have shown that a search for the LFV decays 
$\phi^{0} \rightarrow \tau^\pm \ell_i^\mp$ of neutral Higgs bosons 
($\phi^{0} =h,H$ and $A$) can be useful 
to further constrain the LFV Yukawa couplings
at future collider experiments. 
In particular, 
the LFV decays of the lightest Higgs boson 
can be one of the important probes to find the evidence for the 
extended Higgs sector when the SM-like situation would 
be preferred by the data at forthcoming collider experiments.
The branching ratio for $h \to \tau^\pm \mu^\mp$ can be 
larger than $\mathcal{O}(10^{-3})$ 
except for the high $\tan\beta$ region
under the constraints from the current experimental data 
and also from the theoretical requirements. 
At ILC (and in case at LHC), such a size of 
the branching fractions can be tested. 
Therefore, we conclude that the search of LFV in the Higgs 
boson decay at future colliders 
can further constrain the LFV Yukawa couplings 
especially in the relatively small $\tan\beta$ region 
($\tan\beta \lesssim 30 $),  
where rare tau decay data cannot reach.

\vspace{1cm}
\noindent
{\large \it Acknowledgments}

The authors would like to thank Yasuhiro Okada for speculative comments,
Eric Torrence, Paride Paradisi for helpful comments, 
and Eiichi Takasugi, Koichi Matsuda, Tetsuo Shindou for valuable 
discussions. 
S.K was supported, in part, by Grants-in-Aid of the Ministry 
of Education, Culture, Sports, Science and Technology, Government of 
Japan, Grant No. 17043008.
T.O. was supported, in part, by JSPS Research Fellowship for Young
Scientists, No. 15-03693.


\end{document}